\documentclass[12pt]{article}
\usepackage{amsfonts}
\begin{document}
\begin{center}
{\Large\bf Supertwistor formulation for higher dimensional
superstrings}\\[0.5cm]
{\large D.V.~Uvarov\footnote{E-mail: d\_uvarov@hotmail.com, uvarov@kipt.kharkov.ua}}\\[0.4cm]
{\it NSC Kharkov Institute of Physics and Technology,}\\ {\it 61108 Kharkov, Ukraine}\\[0.5cm]
\end{center}
\begin{abstract}
Considered is the formulation for the superstring action in 6 and
10 dimensions involving supertwistor variables that appropriately
generalize 4-dimensional Ferber supertwistors. Equations of motion
and $\kappa-$symmetry transformations in terms of the
supertwistors are derived. It is shown that covariant
$\kappa-$symmetry gauge fixing reduces superstring action to the
quadratic one with respect to supertwistors.  Upon gauge fixing
remaining symmetries it can be cast into the form of the
light-cone gauge Green-Schwarz superstring action.

\end{abstract}

\section{Introduction}

Twistor theory \cite{twistor} and its supersymmetric
generalization in 4 dimensions \cite{Ferber} are known to unveil
important features of the space-time structure and geometry and
provide valuable insights into the field theory. However, until
recently twistor methods application to string theory was
basically limited to point-like (super)particles
\cite{Shirafuji}-\cite{BdAS}  and the tensionless extended objects
\cite{Ilyenko}-\cite{BdAM} (see, however,
\cite{Shaw}-\cite{FedLuk} for the approaches to the twistor
description of tensile string and brane models). Active
exploration of the gauge fields/strings correspondence using
spinor and twistor methods had begun after Witten's construction
of the topological string in the projective supertwistor space
$\mathbb{CP}^{(3|4)}$ \cite{Witten}. Since then there have been
proposed other twistor superstring models
\cite{Berkovits}-\cite{Bars} sharing the feature that all of them
appear to be different from the conventional Green-Schwarz
superstring both classically and quantum mechanically. This raises
the question of examining the possibility of the twistor
reformulation for the Green-Schwarz superstring and looking for
adequate supertwistor variables. Such twistor transform for the
Green-Schwarz superstring could provide insights into the
covariant quantization problem, as well as, underline the
similarities and differences as compared to known twistor string
models.

In \cite{CQG06} there have been started such investigation by
performing the supertwistor transform for the $D=4$ superstring
using as the starting point the first-order action functional
\cite{BZstring} that involves Lorentz-harmonics
\cite{GIKOS}-\cite{harmonics} as auxiliary variables and on the
classical level is equivalent to the Green-Schwarz formulation. In
terms of Ferber supertwistors \cite{Ferber} was obtained the
string action that, unlike the previously studied formulations for
point-like and extended objects in terms of twistor variables
\cite{Shirafuji}-\cite{PHPW}, \cite{balu}-\cite{BAPV}, is
invariant under irreducible $\kappa-$symmetry transformations. It
was shown that covariant
 $\kappa-$symmetry gauge fixing results in the superstring action
quadratic in the (super)twistor variables that is of the $2d$ free
field theory type modulo the constraints on (super)twistors
necessary to ensure their incidence to the superspace with real
body. The classification of the constraints and structure of the
gauge symmetries such gauge-fixed action is characterized by has
been also considered \cite{U06}.

Present paper is devoted to the twistor transform of the
superstring action in higher dimensions. It requires the proper
generalization of supertwistors to dimensions $D>4$. Such
generalization by itself appears to be rather complicated and
ambiguous as was explored earlier from the geometrical and
field-theoretical perspectives \cite{Hughston}-\cite{Cherkis}, as
well as, from the twistor description of superparticles and
superstrings \cite{balu}, \cite{BAPV}, \cite{Bars}, \cite{Bars2},
\cite{BdAM}. For our consideration we use those $D=6$ and $D=10$
supertwistors, whose projectional part is identified with the
spinor Lorentz-harmonic matrix that allows to establish the
relation with the string momentum density represented by the
Lorentz vector.

Sections 2 and 3 are devoted to the twistor transform for the
$D=6$ superstring. After defining appropriate twistor variables we
present the supertwistor formulation for the superstring action,
derive corresponding equations of motion, adduce $\kappa-$symmetry
transformation rules and consider Lorentz-covariant
$\kappa-$symmetry gauge leading to the simplification of the
superstring action. Generalization of the above results to the
$D=10$ case is the subject of sections 4 and 5. Besides that
section 5 contains discussion of how the $\kappa-$symmetry
gauge-fixed action can be related to the light-cone gauge
formulation of the Green-Schwarz superstring.

\section{Lorentz harmonics and $D=6$ supertwistors}

As the $D=6$ $N=1$ superconformal group is isomorphic to the
$OSp(8^*|2)$ supergroup \cite{CKvP} we consider the supertwistor
to realize its fundamental representation\footnote{Similar to our
approach to the construction of $D=6$ (super)twistors was pursued
in \cite{Bars}, \cite{Bars2}.}
\begin{equation}\label{twistor6}
\mathcal Z^{\Lambda\mathrm a}=( \mu^{\underline\alpha\mathrm a},\
v^{\mathrm a}_{\underline\alpha},\ \eta^{i\mathrm a}).
\end{equation}
Following the Penrose's nomenclature \cite{twistor} we call
$\mu^{\underline\alpha\mathrm a}$ the primary spinor part of
supertwistor and $v^{\mathrm a}_{\underline\alpha}$ its
projectional part. They are represented by the pair of $D=6$
symplectic Majorana-Weyl spinors of opposite chiralities
\begin{equation}
(\mu^{\underline\alpha \mathrm
a})^*=C^{\underline{\dot\alpha}}{}_{\underline\beta}\varepsilon_{\mathrm
a\mathrm b}\mu^{\underline\beta\mathrm b},\quad (v^{\mathrm
a}_{\underline\alpha})^*=C^{-1\underline\beta}{}_{\underline{\dot\alpha}}\varepsilon_{\mathrm
a\mathrm b}v_{\underline\beta}^{\mathrm b},
\end{equation}
where $D=6$ charge conjugation matrix
$C^{\underline{\dot\alpha}}{}_{\underline\beta}$ and its inverse
$C^{-1\underline\beta}{}_{\underline{\dot\alpha}}$ can be
expressed as
\begin{equation}
C^{\underline{\dot\alpha}}{}_{\underline\beta}=\left(
\begin{array}{cccc}
0&1&0&0\\
-1&0&0&0\\
0&0&0&1\\
0&0&-1&0
\end{array}\right),\quad
C^{-1\underline\beta}{}_{\underline{\dot\alpha}}=\left(
\begin{array}{cccc}
0&-1&0&0\\
1&0&0&0\\
0&0&0&-1\\
0&0&1&0
\end{array}\right),\quad
(C^{\underline{\dot\alpha}}{}_{\underline{\beta}})^*=-C^{-1\underline\alpha}{}_{\underline{\dot\beta}}.
\end{equation}
$\varepsilon_{\mathrm a\mathrm b}$ and $\varepsilon^{\mathrm
a\mathrm b}$: $\varepsilon_{\mathrm a\mathrm
b}\varepsilon^{\mathrm b\mathrm c}=\delta_{\mathrm a}^{\mathrm c}$
are unit antisymmetric metric and its inverse in the fundamental
representation of $SU(2)$ that will be identified below with the
constituents of the $SO(4)=SU(2)\times\widetilde{SU(2)}$ group of
rotations in the directions orthogonal to the string world-sheet.
The supertwistor components are assumed to satisfy the incidence
relations
\begin{equation}\label{inc6}
\mu^{\underline\alpha \mathrm a}=v_{\underline\beta}^{\mathrm
a}(x^{\underline{\beta\alpha}}-2i\theta^{\underline\beta}_{i}\theta^{\underline\alpha
i}),\quad\eta^{i\mathrm a}=2v_{\underline\alpha}^{\mathrm
a}\theta^{\underline\alpha i}
\end{equation}
with respect to the $D=6$ $N=1$ superspace coordinates
$x^{\underline{\alpha\beta}}=x^{\underline
m}\tilde\gamma^{\underline{\alpha\beta}}_{\underline m}$,
$\theta^{\underline\alpha i}$, where
$\tilde\gamma^{\underline{\alpha\beta}}_{\underline m}$
($\underline m=0,1,...,5$) are $6d$ chiral $\gamma-$matrices
antisymmetric in spinor indices. Superspace Grassmann coordinates
$\theta^{\underline\alpha i}$ also obey symplectic Majorana-Weyl
condition
\begin{equation}
(\theta^{\underline\alpha
i})^*=-C^{\underline{\dot\alpha}}{}_{\underline\beta}\varepsilon_{
ij}\theta^{\underline\beta j},
\end{equation}
where $\varepsilon_{ij}$ and $\varepsilon^{ij}$ are the metric
tensor and its inverse for the $SU(2)$ $R-$symmetry subgroup of
$OSp(8^*|2)$.

Being interested in the twistor description of $D=6$ superstrings
we identify the projectional part of the supertwistor
(\ref{twistor6}) $v^{\mathrm a}_{\underline\alpha}$ as the
component of the $D=6$ spinor Lorentz harmonic matrix
\begin{equation}\label{spinharm6}
v_{\underline\alpha}^{(\underline\alpha)}=(v_{\underline\alpha}^{+a},
v_{\underline\alpha}^{-\dot a})\in Spin(1,5)
\end{equation}
with the chiral spinor index in brackets realized as the aggregate
of spinor indices corresponding to $SO(1,1)$, to be identified
with the world-sheet structure group, and $SO(4)$. Note that $D=6$
spinor Lorentz harmonics \cite{harmonics} satisfy the reality
\begin{equation}\label{harm6}
(v_{\underline\alpha}^{(\underline\alpha)})^*=-C^{(\underline{\dot\alpha})}{}_{(\underline\beta)}v_{\underline\beta}^{(\underline\beta)}C^{-1\underline\beta}{}_{\underline{\dot\alpha}}
\end{equation}
and unimodularity
\begin{equation}\label{harm6'}
\mathrm{det}v^{(\underline\alpha)}_{\underline\alpha}=1
\end{equation}
constraints reducing the number of their independent components to
the dimension of the $Spin(1,5)$ group equal 15. Hence we have the
following pair of supertwistors associated to $D=6$ superstring
\begin{equation}\label{twistor6str}
\mathcal Z^{\Lambda+a}=(\mu^{\underline\alpha+a},
v_{\underline\alpha}^{+a}, \eta^{i+a}),\quad\mathcal
Z^{\Lambda-\dot a}=(\mu^{\underline\alpha-\dot a},
v_{\underline\alpha}^{-\dot a}, \eta^{i-\dot a}).
\end{equation}
In order to satisfy incidence relations (\ref{inc6}) with
$v_{\underline\alpha}^{\mathrm a}$ standing for either
$v_{\underline\alpha}^{+a}$ or $v_{\underline\alpha}^{-\dot a}$,
supertwistors (\ref{twistor6str}) have to be constrained by 10
relations
\begin{equation}\label{twconstr6}
\mathcal Z^{\Lambda+a}G_{\Lambda\Sigma}\mathcal
Z^{\Sigma+b}=\mathcal Z^{\Lambda-\dot a}G_{\Lambda\Sigma}\mathcal
Z^{\Sigma-\dot b}=\mathcal Z^{\Lambda+a}G_{\Lambda\Sigma}\mathcal
Z^{\Sigma-\dot a}=0
\end{equation}
ensuring elimination of the contribution of the 3-form coordinates
$y^{\underline{\alpha\beta}}=y^{\underline{lmn}}\tilde\gamma^{\underline{\alpha\beta}}_{\underline{lmn}}$.
In (\ref{twconstr6}) invariant scalar product of  supertwistors is
arranged using the $OSp(8^*|2)$ metric
\begin{equation}
G_{\Lambda\Sigma}=\left(
\begin{array}{ccc}
0&\delta_{\underline\alpha}^{\underline\beta} &0\\[0.2cm]
\delta^{\underline\alpha}_{\underline\beta} &0&0\\[0.2cm]
0&0& -i\varepsilon_{ij}
\end{array}\right).
\end{equation}

It is also worthwhile to mention that the $OSp(8^*|2)$
superalgebra can be viewed as the conformal superalgebra
associated to the centrally extended $D=4$ $N=2$ superPoincare
algebra. Corresponding representation for the primary spinor and
projectional parts of the supertwistor (\ref{twistor6})
\begin{equation}
\begin{array}{c}
\mu^{\underline\alpha\mathrm a}=\left(
\mu^{\alpha\mathrm a},\
\bar\mu^{\mathrm a}_{\dot\alpha}
\right),\\[0.2cm]
v_{\underline\alpha}^{\mathrm a} =\left(
v_{\alpha}^{\mathrm a},\
\bar v^{\dot\alpha\mathrm a}
\right)
\end{array}
\end{equation}
is given in terms of two pairs of $SL(2,\mathbb C)$ 2-component
spinors. The incidence relations become
\begin{equation}
\begin{array}{c}
\mu^{\alpha\mathrm a}=\bar v^{\mathrm a}_{\dot\beta}(x^{\dot\beta\alpha}+4i\theta^\alpha_{i}\bar\theta^{\dot\beta i})-v^{\beta\mathrm a}(\bar z\delta^\alpha_\beta+4i\theta^i_{\beta}\theta_i^{\alpha}),\\[0.2cm]
\bar\mu^{\mathrm a}_{\dot\alpha}=-v^{\beta\mathrm a}(x_{\beta\dot\alpha}+4i\theta^i_{\beta}\bar\theta_{\dot\alpha i})+\bar v_{\dot\beta}^{\mathrm a}(z\delta^{\dot\beta}_{\dot\alpha}+4i\bar\theta^{\dot\beta}_{i}\bar\theta^{i}_{\dot\alpha})\\[0.2cm]
\eta^{i\mathrm a}=4(v_\alpha^{\mathrm a}\theta^{\alpha i}+\bar
v^{\dot\alpha\mathrm a}\bar\theta^i_{\dot\alpha}),
\end{array}
\end{equation}
where it has been performed the $4+2$ splitting of $D=6$ Minkowski
coordinates $x^{\underline m}=(x^m,x^4,x^5)$ with
$x_{\alpha\dot\beta}=\sigma_{m\alpha\dot\beta}x^m$ ($m=0,1,2,3$)
and the complex scalar coordinate $z=x^5+ix^4$ parametrizing
bosonic body of the centrally extended $D=4$ $N=2$ superspace.

\section{Supertwistor formulation of $D=6$ superstring}

Appropriate starting point for our consideration is the $D=6$
$N=1$ superstring in the Lorentz-harmonic formulation, whose
action equals
\begin{equation}\label{ssa}
S_{LH}^{D=6}=S_{kin}+S_{WZ},
\end{equation}
where the kinetic term is defined as
\begin{equation}
S_{kin}=\frac{1}{2(\alpha')^{1/2}}\int d^2\xi (e^{+2}n_{\underline
m}^{-2}-e^{-2}n_{\underline m}^{+2})\wedge\omega^{\underline
m}(d)+\frac{c}{2}\int d^2\xi e^{-2}\wedge e^{+2}
\end{equation}
and the Wess-Zumino term is given by the expression
\begin{equation}
S_{WZ}=\frac{is}{c\alpha'}\int d^2\xi\omega^{\underline
m}(d)\wedge d\theta^{\underline\alpha}_{i}\gamma_{\underline
m\underline{\alpha\beta}}\theta^{\underline\beta i}.
\end{equation}
In the superstring action (\ref{ssa}) $\omega^{\underline m}(d)=d
x^{\underline m}-id\theta^{\underline\alpha}_{
i}\gamma_{\underline
m\underline{\alpha\beta}}\theta^{\underline\beta i}$ is the $D=6$
$N=1$ supersymmetric 1-form, $e^{\pm2}_\mu$ ($\mu=\tau,\sigma$)
are the components of the world-sheet zweibein written in the
light-cone basis for the $SO(1,1)$ world-sheet structure group,
$\alpha'$ is the constant of dimension $[L]^2$, $c$ is
dimensionless numerical constant, as well as, $s=\pm1$ both values
of which are consistent with the $\kappa$-invariance of the
action. Action (\ref{ssa}) also contains two light-like vectors
$n_{\underline m}^{\pm2}(\xi)$ from the string adopted $D=6$
vector harmonic matrix $n_{\underline m}^{(\underline
n)}=(n^{+2}_{\underline m}, n^{-2}_{\underline m},
n^{\bar\imath}_{\underline m})$
\begin{equation}\label{norm6}
\begin{array}{c} n_{\underline m}^{(\underline n)}n^{\underline
m(\underline k)}=\eta^{(\underline n)(\underline k)}:\\[0.2cm]
n_{\underline m}^{+2}n^{\underline m+2}=n_{\underline
m}^{-2}n^{\underline m-2}=0,\quad n_{\underline
m}^{+2}n^{\underline m-2}=2,\quad n_{\underline
m}^{\pm2}n^{\underline m\bar\imath}=0,\quad n_{\underline
m}^{\bar\imath}n^{\underline
m\bar\jmath}=-\delta^{\bar\imath\bar\jmath}
\end{array}
\end{equation}
admitting the following realization through the above introduced
spinor harmonics (\ref{spinharm6})
\begin{equation}
n_{\underline
m}^{+2}=\frac12v_{\underline\alpha}^{+a}\tilde\gamma_{\underline
m}^{\underline{\alpha\beta}}v_{\underline\beta}^{+b}\varepsilon_{ab},\quad
n_{\underline m}^{-2}=-\frac12v_{\underline\alpha}^{-\dot
a}\tilde\gamma_{\underline
m}^{\underline{\alpha\beta}}v_{\underline\beta}^{-\dot
b}\varepsilon_{\dot a\dot b},\quad n_{\underline
m}^{\bar\imath}=-\frac12v_{\underline\alpha}^{+a}\tilde\gamma_{\underline
m}^{\underline{\alpha\beta}}v_{\underline\beta}^{-\dot
b}\sigma^{\bar\imath}_{a\dot b},
\end{equation}
where the index $\bar\imath=1,...,4$ belongs to the vector
representation of the $SO(4)$ and $\sigma^{\bar\imath}_{a\dot b}$
are the corresponding $\sigma-$matrices. Differentials/variations
of the vector harmonics consistent with the normalization
relations (\ref{norm6})
\begin{equation}
dn^{\pm2}_{\underline m}=\mp\frac12\Omega^{+2-2}(d)n_{\underline
m}^{\pm2}+\Omega^{\pm2\bar\imath}(d)n^{\bar\imath}_{\underline
m},\quad dn^{\bar\imath}_{\underline
m}=\frac12\Omega^{+2\bar\imath}(d)n^{-2}_{\underline
m}+\frac12\Omega^{-2\bar\imath}(d)n^{+2}_{\underline
m}+\Omega^{\bar\imath\bar\jmath}(d)n_{\underline m}^{\bar\jmath}
\end{equation}
are expressed through the $SO(1,1)\times SO(4)$ split components
of the Cartan 1-form $\Omega^{(\underline k)(\underline l)}(d)=
\frac12(n_{\underline m}^{(\underline k)}dn^{\underline
m(\underline l)}-n_{\underline m}^{(\underline l)}dn^{\underline
m(\underline k)})$:
\begin{equation}\label{cartan6}
\begin{array}{c}
\Omega^{+2-2}(d)=\frac12(n^{+2}_{\underline m}dn^{\underline
m-2}-n^{-2}_{\underline m}dn^{\underline m+2}),\quad
\Omega^{\pm2\bar\imath}(d)=\frac12(n_{\underline
m}^{\pm2}dn^{\underline
m\bar\imath}-n_{\underline m}^{\bar\imath}dn^{\underline m\pm2}),\\[0.2cm]
\Omega^{\bar\imath\bar\jmath}(d)=\frac12(n_{\underline
m}^{\bar\imath}dn^{\underline m\bar\jmath}-n_{\underline
m}^{\bar\jmath}dn^{\underline m\bar\imath})
\end{array}
\end{equation}
invariant under the $SO(1,5)$ transformations acting on indices
without brackets. They are used to derive the equations of motion
for the $D=6$ Lorentz-harmonic superstring among which is the
following nondynamical equation
\begin{equation}\label{rheotr6}
\omega^{\underline
m}(d)={\textstyle\frac{c(\alpha')^{1/2}}{2}}(e^{+2}n^{\underline
m-2}+e^{-2}n^{\underline m+2})
\end{equation}
that defines the positioning w.r.t. to the world-sheet of the
supersymmetric 1-form $\omega^{\underline m}(d)$. Note that
light-like vectors $n^{\pm2}_{\underline m}$ can be identified
with the world-sheet tangents, while other components of the
harmonic matrix $n^{\bar\imath}_{\underline m}$ are orthogonal to
the world-sheet. Relation (\ref{rheotr6}) is used to recover the
Green-Schwarz action from the Lorentz-harmonic one (\ref{ssa})
\begin{equation}
S^{D=6}_{GS}=-\frac{1}{2c\alpha'}\int
d^2\xi\sqrt{-g}g^{\mu\nu}\omega^{\underline
m}_\mu\omega_{\underline m\nu}+\frac{is}{c\alpha'}\int
d^2\xi\varepsilon^{\mu\nu}\omega^{\underline
m}_\mu\partial_\nu\theta_{i}\gamma_{\underline m}\theta^{i},
\end{equation}
where the inverse of the world-sheet metric
$g^{\mu\nu}=\frac12(e^{\mu+2}e^{\nu-2}+e^{\mu-2}e^{\mu-2})$ is
defined by the components of the inverse zweibein  $e^{\mu\pm2}$
and $\sqrt{-g}=e=\mathrm{det}(e^{+2}_\mu,e^{-2}_\mu)$, thus
establishing the classical equivalence of both formulations.

Twistorization of the Lorentz-harmonic superstring action
(\ref{ssa}) proceeds by representing projections of the
supersymmetric 1-form $\omega^{\underline m}(d)$ onto the tangent
to the world-sheet vector harmonic components
$n^{\pm2}_{\underline m}$ in terms of the supertwistor variables
(\ref{twistor6str})
\begin{equation}\label{twist6}
\begin{array}{rl}
\omega^{+2}(d)=& n^{+2}_{\underline m}\omega^{\underline
m}(d)=\frac12\varepsilon_{ab}d\mathcal
Z^{\Lambda+a}G_{\Lambda\Sigma}\mathcal Z^{\Sigma+b}\\[0.2cm]
=&\frac12\varepsilon_{ab}(d
\mu^{\underline\alpha+a}v_{\underline\alpha}^{+b}+d
v_{\underline\alpha}^{+a}\mu^{\underline\alpha+b}-i\varepsilon_{ij}d\eta^{i+a}\eta^{j+b}),\\[0.2cm]
\omega^{-2}(d)=& n^{-2}_{\underline m}\omega^{\underline
m}(d)=-\frac12\varepsilon_{\dot a\dot b}d\mathcal Z^{\Lambda-\dot
a}G_{\Lambda\Sigma}\mathcal Z^{\Sigma-\dot b}\\[0.2cm]
=&-\frac12\varepsilon_{\dot a\dot b}(d\mu^{\underline\alpha-\dot
a}v_{\underline\alpha}^{-\dot b}+dv_{\underline\alpha}^{-\dot
a}\mu^{\underline\alpha-\dot b}-i\varepsilon_{ij}d\eta^{i-\dot
a}\eta^{j-\dot b}).
\end{array}
\end{equation}
Quite analogously can be expressed in terms of supertwistors
$\omega^{\underline m}(d)$ projections onto the orthogonal to the
world-sheet components of the vector harmonic matrix
\begin{equation}\label{twist6'}
\omega^{\bar\imath}(d)=n^{\bar\imath}_{\underline
m}\omega^{\underline m}(d)=\frac14(\mathcal
Z^{\Lambda+a}G_{\Lambda\Sigma}d\mathcal Z^{\Sigma-\dot
a}-d\mathcal Z^{\Lambda+a}G_{\Lambda\Sigma}\mathcal Z^{\Sigma-\dot
a})\sigma^{\bar\imath}_{a\dot a}.
\end{equation}
Then the first-order $D=6$ superstring action (\ref{ssa}) is
represented as
\begin{equation}\label{ssa-tw}
\begin{array}{c}
S^{D=6}_{tw}=\frac{1}{2(\alpha')^{1/2}}\int d^2\xi(e^{+2}\wedge\omega^{-2}(d)-e^{-2}\wedge\omega^{+2}(d))+\frac{c}{2}\int d^2\xi e^{-2}\wedge e^{+2}\\[0.2cm]
+\frac{is}{c\alpha'}\int
d^2\xi(\frac12\omega^{-2}(d)\wedge\varphi^{+2}(d)+\frac12\omega^{+2}(d)\wedge\varphi^{-2}(d)-\omega^{\bar\imath}(d)\wedge\varphi^{\bar\imath}(d)),
\end{array}
\end{equation}
where additionally the following 1-forms quadratic in the
Grassmann-odd supertwistor components
\begin{equation}
\begin{array}{c}
\varphi^{+2}(d)=-\frac12\varepsilon_{ab}\mathcal
D\eta^{+a}_{i}\eta^{i+b},\quad
\varphi^{-2}(d)=\frac12\varepsilon_{\dot a\dot b}\mathcal
D\eta^{-\dot a}_{i}\eta^{i-\dot b},
\\[0.2cm]
\varphi^{\bar\imath}(d)=\frac14(\mathcal
D\eta^{+a}_{i}\eta^{i-\dot a}-\mathcal D\eta^{-\dot
a}_{i}\eta^{i+a})\sigma^{\bar\imath}_{a\dot a}
\end{array}
\end{equation}
have been introduced. Covariant differentials for the
Grassmann-odd supertwistor components that enter above expressions
equal
\begin{equation}
\begin{array}{c}
\mathcal
D\eta^{+a}_{i}=d\eta^{+a}_{i}+\frac14\Omega^{+2-2}(d)\eta_{i}^{+a}-\frac12\Omega^{+2\bar\imath}(d)\tilde\sigma^{\bar\imath\dot
aa}\eta_{i\dot a}^-
-\frac12\Omega^{\bar\imath\bar\jmath}(d)\sigma^{\bar\imath\bar\jmath}{}_{b}{}^a\eta_{i}^{+b},\\[0.2cm]
\mathcal D\eta^{-\dot a}_{i}=d\eta^{-\dot
a}_{i}-\frac14\Omega^{+2-2}(d)\eta_{i}^{-\dot
a}-\frac12\Omega^{-2\bar\imath}(d)\tilde\sigma^{\bar\imath\dot
aa}\eta_{i
a}^++\frac12\Omega^{\bar\imath\bar\jmath}(d)\tilde\sigma^{\bar\imath\bar\jmath}{}^{\dot
a}{}_{\dot b}\eta_{i}^{-\dot b},
\end{array}
\end{equation}
where $\tilde\sigma^{\bar\imath\bar\jmath}{}^{\dot a}{}_{\dot
b}=\frac14(\tilde\sigma^{\bar\imath\dot
aa}\sigma^{\bar\jmath}_{a\dot b}-\tilde\sigma^{\bar\jmath\dot
aa}\sigma^{\bar\imath}_{a\dot b})$,
$\sigma^{\bar\imath\bar\jmath}{}_{b}{}^{a}=\frac14(\sigma^{\bar\imath}_{b\dot
a}\tilde\sigma^{\bar\jmath\dot aa}-\sigma^{\bar\jmath}_{b\dot
a}\tilde\sigma^{\bar\imath\dot aa})$.

Because of the presence of algebraic constraints (\ref{twconstr6})
admissible differentials/variations for supertwistors should be
introduced that take them into account. Those for spinor Lorentz
harmonics are given by
\begin{equation}
\begin{array}{c}
dv_{\underline\alpha}^{+a}=-\frac14\Omega^{+2-2}(d)v_{\underline\alpha}^{+a}+\frac12\Omega^{+2\bar\imath}(d)\tilde\sigma^{\bar\imath\dot aa}v_{\underline\alpha\dot a}^-+\frac12\Omega^{\bar\imath\bar\jmath}(d)\sigma^{\bar\imath\bar\jmath}{}_{b}{}^av_{\underline\alpha}^{+b},\\[0.2cm]
dv_{\underline\alpha}^{-\dot
a}=\frac14\Omega^{+2-2}(d)v_{\underline\alpha}^{-\dot
a}+\frac12\Omega^{-2\bar\imath}(d)\tilde\sigma^{\bar\imath\dot
aa}v_{\underline\alpha
a}^+-\frac12\Omega^{\bar\imath\bar\jmath}(d)\tilde\sigma^{\bar\imath\bar\jmath}{}^{\dot
a}{}_{\dot b}v_{\underline\alpha}^{-\dot b}.
\end{array}
\end{equation}
Derivation coefficients (\ref{cartan6}) can equivalently be
written in terms of the spinor harmonics as follows
\begin{equation}\label{cartan6s}
\begin{array}{c}
\Omega^{+2-2}(d)=v^{\underline\alpha+}_{\dot a}dv^{-\dot a}_{\underline\alpha}-v^{\underline\alpha-}_adv^{+a}_{\underline\alpha},\quad\Omega^{+2\bar\imath}(d)=v^{\underline\alpha+\dot a}dv^{+a}_{\underline\alpha}\sigma^{\bar\imath}_{a\dot a},\quad\Omega^{-2\bar\imath}(d)=v^{\underline\alpha-a}dv_{\underline\alpha}^{-\dot a}\sigma^{\bar\imath}_{a\dot a},\\[0.2cm]
\Omega^{\bar\imath\bar\jmath}(d)=v^{\underline\alpha+}_{\dot
a}dv_{\underline\alpha}^{-\dot
b}\tilde\sigma^{\bar\imath\bar\jmath}{}^{\dot a}{}_{\dot
b}-v^{\underline\alpha-}_{a}dv_{\underline\alpha}^{+b}\sigma^{\bar\imath\bar\jmath}{}_{b}{}^{a}.
\end{array}
\end{equation}
In (\ref{cartan6s}) there are present also the components of the
inverse spinor harmonic matrix
\begin{equation}
v^{\underline\alpha}_{(\underline\alpha)}=(v^{\underline\alpha-}_a,
v^{\underline\alpha+}_{\dot a}):\quad
v^{(\underline\beta)}_{\underline\alpha}v^{\underline\alpha}_{(\underline\alpha)}=
\delta^{(\underline\beta)}_{(\underline\alpha)}
\end{equation}
that can either be considered as independent variables that obey
the above constraints or be directly expressed via the spinor
harmonics as
\begin{equation}
v^{\underline\alpha}_{(\underline\alpha)}=\frac16\varepsilon^{\underline{\alpha\beta\gamma\delta}}\varepsilon_{(\underline\alpha)(\underline\beta)(\underline\gamma)(\underline\delta)}v_{\underline\beta}^{(\underline\beta)}v_{\underline\gamma}^{(\underline\gamma)}v^{(\underline\delta)}_{\underline\delta}.
\end{equation}
Then the expressions for the admissible differentials of
supertwistors respecting the constraints (\ref{twconstr6}) read
\begin{equation}
\begin{array}{rl}
d\mathcal Z^{\Lambda+a}=&-\frac14\Omega^{+2-2}(d)\mathcal
Z^{\Lambda+a}+\frac12\Omega^{+2\bar\imath}(d)\tilde\sigma^{\bar\imath\dot
aa}\mathcal Z^{\Lambda-}_{\dot
a}+\frac12\Omega^{\bar\imath\bar\jmath}(d)\sigma^{\bar\imath\bar\jmath}{}_{b}{}^a\mathcal Z^{\Lambda+b}\\[0.2cm]
&-(\omega^{+2}(d)\varepsilon^{ab}+i\mathcal
D\eta^{+a}_{i}\eta^{i+b})V^{\Lambda-}_b+(\omega^{\bar\imath}(d)\tilde\sigma^{\bar\imath\dot
aa}-i\mathcal D\eta^{+a}_{i}\eta^{i-\dot
a})V^{\Lambda+}_{\dot a}+J^\Lambda_\Sigma\mathcal D\mathcal Z^{\Sigma+a}\\[0.2cm]
d\mathcal Z^{\Lambda-\dot a}=&\frac14\Omega^{+2-2}(d)\mathcal
Z^{\Lambda-\dot
a}+\frac12\Omega^{-2\bar\imath}(d)\tilde\sigma^{\bar\imath\dot
aa}\mathcal Z^{\Lambda+}_{
a}-\frac12\Omega^{\bar\imath\bar\jmath}(d)\tilde\sigma^{\bar\imath\bar\jmath}{}^{\dot
a}{}_{\dot
b}\mathcal Z^{\Lambda-\dot b}\\[0.2cm]
&+(\omega^{-2}(d)\varepsilon^{\dot a\dot b}-i\mathcal D\eta^{-\dot
a}_{i}\eta^{i-\dot b})V^{\Lambda+}_{\dot
b}-(\omega^{\bar\imath}(d)\tilde\sigma^{\bar\imath\dot
aa}+i\mathcal D\eta^{-\dot
a}_{i}\eta^{i+a})V^{\Lambda-}_{a}+J^\Lambda_\Sigma\mathcal
D\mathcal Z^{\Sigma-\dot a},
\end{array}
\end{equation}
where
\begin{equation}
V^{\Lambda-}_{a}=(v^{\underline\alpha-}_a,0,0),\quad
V^{\Lambda+}_{\dot a}=(v^{\underline\alpha+}_{\dot a},0,0)
\end{equation}
are the supertwistors associated to inverse spinor harmonics
$v^{\underline\alpha-}_a$, $v^{\underline\alpha+}_{\dot a}$ and
the matrix
\begin{equation}
J^\Lambda_\Sigma=\left(
\begin{array}{ccc}
0&0&0\\
0&0&0\\
0&0&\delta^i_j
\end{array}\right)
\end{equation}
singles out Grassmann-odd components of the supertwistors.

To derive equations of motion for the superstring in the
supertwistor formulation we note that the differentials of 1-forms
entering the action (\ref{ssa-tw}) equal
\begin{equation}
\begin{array}{rl}
d\omega^{+2}=&\frac12\Omega^{+2-2}(d)\wedge\omega^{+2}(d)-\Omega^{+2\bar\imath}(d)\wedge\omega^{\bar\imath}(d)+\frac{i}{2}\mathcal D\eta^{+a}_{i}\wedge\mathcal D\eta^{i+}_{a},\\[0.2cm]
d\omega^{-2}=&-\frac12\Omega^{+2-2}(d)\wedge\omega^{-2}(d)-\Omega^{-2\bar\imath}(d)\wedge\omega^{\bar\imath}(d)-\frac{i}{2}\mathcal D\eta^{-\dot a}_{i}\wedge\mathcal D\eta^{i-}_{\dot a},\\[0.2cm]
d\omega^{\bar\imath}=&-\frac12\Omega^{+2\bar\imath}(d)\wedge\omega^{-2}(d)-\frac12\Omega^{-2\bar\imath}(d)\wedge\omega^{+2}(d)-\Omega^{\bar\imath\bar\jmath}(d)\wedge\omega^{\bar\jmath}(d)\\[0.2cm]
+&\frac{i}{2}\mathcal D\eta^{-\dot a}_{i}\wedge\mathcal
D\eta^{i+a}\sigma^{\bar\imath}_{a\dot a}
\end{array}
\end{equation}
and
\begin{equation}
\begin{array}{rl}
d\varphi^{+2}=&\frac12\Omega^{+2-2}(d)\wedge\varphi^{+2}(d)-\Omega^{+2\bar\imath}(d)\wedge\varphi^{\bar\imath}(d)-\frac{1}{2}\mathcal D\eta^{+a}_{i}\wedge\mathcal D\eta^{i+}_{a},\\[0.2cm]
d\varphi^{-2}=&-\frac12\Omega^{+2-2}(d)\wedge\varphi^{-2}(d)-\Omega^{-2\bar\imath}(d)\wedge\varphi^{\bar\imath}(d)+\frac{1}{2}\mathcal D\eta^{-\dot a}_{i}\wedge\mathcal D\eta^{i-}_{\dot a},\\[0.2cm]
d\varphi^{\bar\imath}=&-\frac12\Omega^{+2\bar\imath}(d)\wedge\varphi^{-2}(d)-\frac12\Omega^{-2\bar\imath}(d)\wedge\varphi^{+2}(d)-\Omega^{\bar\imath\bar\jmath}(d)\wedge\varphi^{\bar\jmath}(d)\\[0.2cm]
+&\frac{1}{2}\mathcal D\eta^{+a}_{i}\wedge\mathcal D\eta^{i-\dot
a}\sigma^{\bar\imath}_{a\dot a}.
\end{array}
\end{equation}
Then the variation of the superstring action (\ref{ssa-tw}) can be
presented as
\begin{equation}
\begin{array}{rl}
\delta S^{D=6}_{tw}=&\frac{1}{2(\alpha')^{1/2}}\int d^2\xi\left(e^{+2}\wedge\left[-\frac12\Omega^{+2-2}(d)\omega^{-2}(\delta)+\frac12\Omega^{+2-2}(\delta)\omega^{-2}(d)-\Omega^{-2\bar\imath}(d)\omega^{\bar\imath}(\delta)\right.\right.\\[0.2cm]
+&\left.\Omega^{-2\bar\imath}(\delta)\omega^{\bar\imath}(d)-i\mathcal
D\eta^{-\dot a}_{i}\mathcal D(\delta)\eta^{i-}_{\dot
a}\right]-de^{+2}\omega^{-2}(\delta)+\delta
e^{+2}\wedge\omega^{-2}(d)\\[0.2cm]
-&
e^{-2}\wedge\left[\frac12\Omega^{+2-2}(d)\omega^{+2}(\delta)-\frac12\Omega^{+2-2}(\delta)\omega^{+2}(d)
-\Omega^{+2\bar\imath}(d)\omega^{\bar\imath}(\delta)\right.\\[0.2cm]
+&\left.\Omega^{+2\bar\imath}(\delta)\omega^{\bar\imath}(d)+i\mathcal
D\eta^{+a}_{i}\mathcal
D(\delta)\eta^{i+}_{a}\right]+de^{-2}\omega^{+2}(\delta)-\delta
e^{-2}\wedge\omega^{+2}(d)\\[0.2cm]
+&\left.c(\alpha')^{1/2}(\delta e^{-2}\wedge e^{+2}+e^{-2}\wedge\delta e^{+2})\right)\\[0.2cm]
+&\frac{is}{2c\alpha'}\int d^2\xi\left(\omega^{+2}(d)\wedge\mathcal D\eta^{-\dot a}_{i}\mathcal D(\delta)\eta^{i-}_{\dot a}+\frac12\omega^{+2}(\delta)\mathcal D\eta^{-\dot a}_{i}\wedge\mathcal D\eta^{i-}_{\dot a}\right.\\[0.2cm]
-&\omega^{-2}(d)\wedge\mathcal D\eta^{+a}_{i}\mathcal
D(\delta)\eta^{i+}_{a}-\frac12\omega^{-2}(\delta)\mathcal
D\eta^{+a}_{i}\wedge\mathcal
D\eta^{i+}_{a}\\[0.2cm]
-&\left.[\omega^{\bar\imath}(d)\wedge(\mathcal D\eta^{i-\dot
a}\mathcal D(\delta)\eta^{+a}_{i}+\mathcal D\eta^{+a}_{i}\mathcal
D(\delta)\eta^{i-\dot a})+\omega^{\bar\imath}(\delta)\mathcal
D\eta^{+a}_{i}\wedge\mathcal D\eta^{i-\dot
a}]\sigma^{\bar\imath}_{a\dot a}\right).
\end{array}
\end{equation}
Equations of motion following from the nullification of the
variation w.r.t. zweibein components and
$\Omega^{\pm2\bar\imath}(\delta)$ can be cast into the form
\begin{equation}\label{rheotr6tw}
\omega^{\pm2}(d)=c(\alpha')^{1/2}e^{\pm2},\quad\omega^{\bar\imath}(d)=0
\end{equation}
and be recognized as the supertwistor counterparts of the
nondynamical equations (\ref{rheotr6}). Note that the equations of
motion corresponding to $\Omega^{+2-2}(\delta)$ and
$\Omega^{\bar\imath\bar\jmath}(\delta)$ appear to be trivial
manifesting the $SO(1,1)\times SO(4)$ gauge invariance of the
action functionals (\ref{ssa}), (\ref{ssa-tw}). Nullification of
the variation w.r.t. $\omega^{\pm2}(\delta)$ and
$\omega^{\bar\imath}(\delta)$ yields the set of equations
\begin{equation}
\begin{array}{c}
de^{+2}+\frac12e^{+2}\wedge\Omega^{+2-2}(d)+\frac{is}{2c(\alpha')^{1/2}}\mathcal
D\eta^{+a}_{i}\wedge\mathcal D\eta^{i+}_{a}=0,\\[0.2cm]
de^{-2}-\frac12e^{-2}\wedge\Omega^{+2-2}(d)+\frac{is}{2c(\alpha')^{1/2}}\mathcal
D\eta^{-\dot a}_{i}\wedge\mathcal D\eta^{i-}_{\dot a}=0,\\[0.2cm]
e^{+2}\wedge\Omega^{-2\bar\imath}(d)-e^{-2}\wedge\Omega^{+2\bar\imath}(d)-\frac{is}{c(\alpha')^{1/2}}\mathcal
D\eta^{+a}_{i}\wedge\mathcal D\eta^{i-\dot
a}\sigma^{\bar\imath}_{a\dot a}=0.
\end{array}
\end{equation}
The first two of them are satisfied identically as the consequence
of the reparametrization invariance of the action functional.
Remaining fermionic equations following from the variation w.r.t.
$\mathcal D(\delta)\eta^{+a}_{i}$, $\mathcal D(\delta)\eta^{-\dot
a}_{i}$ then acquire the form
\begin{equation}
(1+s)e^{-2}\wedge\mathcal D\eta^{i+}_a=(1-s)e^{+2}\wedge\mathcal
D\eta^{i-}_{\dot a}=0.
\end{equation}
Definite choice of the numerical factor $s=\pm1$ turns one of the
above equations into identity being the manifestation of the
$\kappa$-symmetry invariance.

Explicit form of the $\kappa-$symmetry transformation rules depend
on the value of $s$. When $s=1$ we have
\begin{equation}
\begin{array}{c}
\delta_\kappa\mathcal Z^{\Lambda+a}=\frac12\Omega^{+2\bar\imath}(\delta_\kappa)\tilde\sigma^{\bar\imath\dot aa}\mathcal Z^{\Lambda-}_{\dot a},\\[0.2cm]
\delta_\kappa\mathcal Z^{\Lambda-\dot
a}=\frac12\Omega^{-2\bar\imath}(\delta_\kappa)\tilde\sigma^{\bar\imath\dot
aa}\mathcal Z^{\Lambda+}_{a}-(K^{\Sigma-\dot a}\mathcal
Z_{\Sigma}^{-\dot b})V^{\Lambda+}_{\dot b}-(K^{\Sigma-\dot
a}\mathcal Z_{\Sigma}^{+ b})V^{\Lambda-}_{b}+K^{\Lambda-\dot
a},\\[0.2cm]
\delta_\kappa e^{+2}=0,\quad\delta_\kappa
e^{-2}=\frac{1}{c(\alpha')^{1/2}}K^{\Lambda-\dot a}\mathcal
D\mathcal Z^-_{\Lambda\dot a},
\end{array}
\end{equation}
where
\begin{equation}
\Omega^{\pm2\bar\imath}(\delta_\kappa)=\pm\frac{1}{c(\alpha')^{1/2}}K^{\Lambda-\dot
b}e^{\nu\pm2}\mathcal D_\nu\mathcal
Z_\Lambda^{+a}\sigma^{\bar\imath}_{a\dot b}
\end{equation}
and the local parameter $\kappa^{i-\dot a}(\xi)$ is presented in
the supertwistor form as $K^{\Lambda-\dot a}=(0,0,\kappa^{i-\dot
a})$. For the $s=-1$ case $\kappa-$symmetry transformations read
\begin{equation}
\begin{array}{c}
\delta_\kappa\mathcal
Z^{\Lambda+a}=\frac12\Omega^{+2\bar\imath}(\delta_\kappa)\tilde\sigma^{\bar\imath\dot
aa}\mathcal Z^{\Lambda-}_{\dot a}-(K^{\Sigma+a}\mathcal
Z_{\Sigma}^{+b})V^{\Lambda-}_b-(K^{\Sigma+a}\mathcal
Z_{\Sigma}^{-\dot b})V^{\Lambda+}_{\dot b}+K^{\Lambda+a},\\[0.2cm]
\delta_\kappa\mathcal Z^{\Lambda-\dot
a}=\frac12\Omega^{-2\bar\imath}(\delta_\kappa)\tilde\sigma^{\bar\imath\dot
aa}\mathcal Z^{\Lambda+}_{a},\\[0.2cm]
\delta_\kappa
e^{+2}=\frac{1}{c(\alpha')^{1/2}}K^{\Lambda+}_{a}\mathcal
D\mathcal Z^{+a}_\Lambda,\quad\delta_\kappa e^{-2}=0,
\end{array}
\end{equation}
where
\begin{equation}
\Omega^{\pm2\bar\imath}(\delta_\kappa)=\pm\frac{1}{c(\alpha')^{1/2}}K^{\Lambda+a}e^{\nu\pm2}\mathcal
D_\nu\mathcal Z^{-\dot b}_\Lambda\sigma^{\bar\imath}_{a\dot b},
\end{equation}
and $K^{\Lambda+a}=(0,0, \kappa^{i+a})$.

Following the lines of discussion in \cite{CQG06} of the
4-dimensional superstring, it is possible to simplify the action
functional (\ref{ssa-tw}) by covariantly gauge fixing
$\kappa-$symmetry. This is can be achieved by substituting
(\ref{rheotr6tw}) into the WZ term
\begin{equation}
S_{WZ}|_{gf}={\textstyle\frac{is}{2(\alpha)^{1/2}}}\int
d^2\xi\left(e^{+2}\wedge\varphi^{-2}(d)+e^{-2}\wedge\varphi^{+2}(d)\right).
\end{equation}
Summing up this expression with that for the kinetic term in
(\ref{ssa-tw}) produces the action functional
\begin{equation}
\begin{array}{rl}
S^{D=6}_{gf}=&{\textstyle\frac{1}{2(\alpha)^{1/2}}}\int
d^2\xi\left[ e^{+2}\wedge(\omega^{-2}(d)+is\varphi^{-2}(d))-
e^{-2}\wedge(\omega^{+2}(d)-is\varphi^{+2}(d))\right]\\[0.2cm]
+&{\textstyle\frac{c}{2}}\int d^2\xi e^{-2}\wedge e^{+2}.
\end{array}
\end{equation}
Further choosing $s=1$ one obtains
\begin{equation}\label{gf6+1}
\begin{array}{rl}
S^{D=6}_{gf,s=1}=&{\textstyle\frac{1}{4(\alpha')^{1/2}}}\int
d^2\xi\left( e^{+2}\wedge dz^{\hat\alpha-}_{\dot a}z^{-\dot
a}_{\hat\alpha}+e^{-2}\wedge d\tilde{\mathcal
Z}^{\Lambda+}_{a}\tilde{\mathcal Z}^{+a}_{\Lambda}\right)\\[0.2cm]
+&\frac{c}{2}\int d^2\xi e^{-2}\wedge e^{+2},
\end{array}
\end{equation}
and correspondingly when $s=-1$
\begin{equation}\label{gf6-1}
\begin{array}{rl}
S^{D=6}_{gf,s=-1}=&{\textstyle\frac{1}{4(\alpha')^{1/2}}}\int
d^2\xi\left( e^{+2}\wedge d\tilde{\mathcal Z}^{\Lambda-}_{\dot
a}\tilde{\mathcal Z}_{\Lambda}^{-\dot a}+e^{-2}\wedge d
z^{\hat\alpha+}_{a}z_{\hat\alpha}^{+a}\right)\\[0.2cm]
+&\frac{c}{2}\int d^2\xi e^{-2}\wedge e^{+2}.
\end{array}
\end{equation}
$z^{\hat\alpha+a}$ and $z^{\hat\alpha-\dot a}$ are bosonic $D=6$
twistors that can be identified as the $Spin(6,2)=Spin(8^*)$
symplectic Majorana-Weyl spinors
\begin{equation}
(z^{\hat\alpha+a})^*=B^{\hat\alpha}{}_{\hat\beta}\varepsilon_{ab}z^{\hat\beta+b},\quad
(z^{\hat\alpha-\dot
a})^*=B^{\hat\alpha}{}_{\hat\beta}\varepsilon_{\dot a\dot
b}z^{\hat\beta-\dot b}.
\end{equation}
In (\ref{gf6+1}), (\ref{gf6-1}) has been also assumed the
following redefinition of the supertwistor components
\begin{equation}\label{tilde6}
\begin{array}{c}
\tilde{\mathcal
Z}^{\Lambda+a}=(\tilde\mu^{\underline\alpha+a}, v^{+a}_{\underline\alpha}, \tilde\eta^{i+a}):\quad\tilde\mu^{\underline\alpha+a}=v_{\underline\beta}^{+a}(x^{\underline{\beta\alpha}}-4i\theta^{\underline\beta}_{i}\theta^{\underline\alpha i}),\quad\tilde\eta^{i+a}=4v_{\underline\alpha}^{+a}\theta^{\underline\alpha i},\\
\tilde{\mathcal Z}^{\Lambda-\dot
a}=(\tilde\mu^{\underline\alpha-\dot a}, v^{-\dot
a}_{\underline\alpha}, \tilde\eta^{i-\dot
a}):\quad\tilde\mu^{\underline\alpha-\dot
a}=v_{\underline\beta}^{-\dot
a}(x^{\underline{\beta\alpha}}-4i\theta^{\underline\beta}_{i}\theta^{\underline\alpha
i}),\quad\tilde\eta^{i-\dot a}=4v_{\underline\alpha}^{-\dot
a}\theta^{\underline\alpha i}
\end{array}
\end{equation}
and the $OSp(8^*|2)$ metric
\begin{equation}
\tilde G_{\Lambda\Sigma}=\left(
\begin{array}{ccc}
0&\delta_{\underline\alpha}^{\underline\beta} &0\\[0.2cm]
\delta^{\underline\alpha}_{\underline\beta} &0&0\\[0.2cm]
0&0& -\frac{i}{2}\varepsilon_{ij}
\end{array}\right).
\end{equation}

It is readily possible to extend the above results for the case of
$D=6$ $N=(2,0)$ superstring. Following the same reasoning and
choosing the $\kappa-$symmetry gauge as $\eta^{1i+a}=\eta^{2i+a}$,
$\eta^{1i-\dot a}=\eta^{2i-\dot a}$ we arrive at the action
functional
\begin{equation}\label{gf6-2}
\begin{array}{rl}
S^{D=6,N=(2,0)}_{gf}=&{\textstyle\frac{1}{4(\alpha')^{1/2}}}\int
d^2\xi\left( e^{+2}\wedge d\tilde{\mathcal Z}^{\Lambda-}_{\dot
a}\tilde{\mathcal Z}_{\Lambda}^{-\dot a}+e^{-2}\wedge
d\tilde{\mathcal Z}^{\Lambda+}_{a}\tilde{\mathcal
Z}_{\Lambda}^{+a}\right)\\[0.2cm]
+&\frac{c}{2}\int d^2\xi e^{-2}\wedge e^{+2}
\end{array}
\end{equation}
formulated in terms of $N=1$ supertwistors (\ref{tilde6}). It
should be noted that (super)twistors entering gauge fixed actions
(\ref{gf6+1}), (\ref{gf6-1}), (\ref{gf6-2}) satisfy algebraic
constraints analogous to (\ref{twconstr6}) that represent the
mixture of the first-class constraints generating $SU(2)\times
SU(2)$ gauge symmetry and the second-class ones.

\section{Lorentz harmonics and $D=10$ supertwistors}

$D=10$ supertwistor can be defined as realizing the fundamental
representation of the minimal superconformal group (according to
the classification developed in \cite{vhvp}) that contains $D=10$
conformal group and is isomorphic to $OSp(1|32)$
\begin{equation}\label{twistor10}
Z^{\mathbf\Lambda}=(\mu^{\hat{\underline\alpha}},
v_{\hat{\underline\alpha}}, \eta).
\end{equation}
The supertwistor is composed of the pair of $Spin(1,9)$
Majorana-Weyl spinors of opposite chiralities and the
Grassmann-odd scalar. Application to the $D=10$ superstring
description suggests to consider the set of 8 supertwistors
\begin{equation}\label{twistor10str}
Z^{\mathbf\Lambda}_{\rm A}=(\mu^{\hat{\underline\alpha}}_{\rm A},
v_{\hat{\underline\alpha}\rm A}, \eta_{\rm A})
\end{equation}
with the label $\rm A=1,...,8$ corresponding to $8_c$ or $8_s$
representation of the covering of transverse Lorentz group $SO(8)$
in $10d$. Supertwistor components are assumed to be incident to
$D=10$ $N=1$ superspace coordinates $(x^{\hat{\underline m}},
\theta^{\hat{\underline\alpha}})$ as follows
\begin{equation}\label{deftwistor}
\mu^{\hat{\underline\alpha}}_{\rm
A}=(x^{\hat{\underline\alpha}\hat{\underline\beta}}-8i\theta^{\hat{\underline\alpha}}\theta^{\hat{\underline\beta}})v_{\hat{\underline\beta}\rm
A},\quad\eta_{\rm A}=4v_{\hat{\underline\alpha}\rm
A}\theta^{\hat{\underline\alpha}},
\end{equation}
where
$x^{\hat{\underline\alpha}\hat{\underline\beta}}=x^{\hat{\underline
m}}\tilde\sigma_{\hat{\underline
m}}^{\hat{\underline\alpha}\hat{\underline\beta}}$
($\hat{\underline m}=0,1,...,9$) and symmetric matrices
$\tilde\sigma_{\hat{\underline
m}}^{\hat{\underline\alpha}\hat{\underline\beta}}$,
$\sigma_{\hat{\underline
m}\hat{\underline\alpha}\hat{\underline\beta}}$ represent $D=10$
chiral $\gamma-$matrices.

Following the consideration of section 2 we identify the
projectional part of the supertwistor (\ref{twistor10str}) as the
$D=10$ spinor Lorentz harmonic matrix
\begin{equation}
v_{\hat{\underline\alpha}}^{({\hat{\underline\alpha}})}=(v^+_{\hat{\underline\alpha}
A},v^-_{\hat{\underline\alpha}\dot A})\in Spin(1,9)
\end{equation}
with the spinor index in brackets split according to the spinor
representations of $SO(1,1)\times SO(8)$. Thus we have the
following associated to $D=10$ superstring sets of supertwistor
variables
\begin{equation}\label{deftwist10}
Z^{\mathbf\Lambda+}_A=(\mu^{\hat{\underline\alpha}+}_A,
v^+_{\hat{\underline\alpha} A}, \eta^+_A),\quad
Z^{\mathbf\Lambda-}_{\dot A}=(\mu^{\hat{\underline\alpha}-}_{\dot
A}, v^-_{\hat{\underline\alpha}\dot A}, \eta^-_{\dot A})
\end{equation}
briefly discussed in \cite{BdAM}. They should obey the constraints
\begin{equation}\label{x3}
Z^{\mathbf\Lambda+}_AG_{\mathbf{\Lambda\Sigma}}Z^{\mathbf\Sigma+}_B=Z^{\mathbf\Lambda-}_{\dot
A}G_{\mathbf{\Lambda\Sigma}}Z^{\mathbf\Sigma-}_{\dot
B}=Z^{\mathbf\Lambda+}_AG_{\mathbf{\Lambda\Sigma}}Z^{\mathbf\Sigma-}_{\dot
A}=0,
\end{equation}
which exclude the contribution to (\ref{deftwistor}) of
$y^{\hat{\underline\alpha}\hat{\underline\beta}}=y^{\hat{\underline
m}\hat{\underline n}\hat{\underline
k}}\tilde\sigma^{\hat{\underline\alpha}\hat{\underline\beta}}_{\hat{\underline
m}\hat{\underline n}\hat{\underline k}}$ coordinates antisymmetric
in spinor indices, as well as,
\begin{equation}\label{x5}
\sigma_{\hat{\underline m}_1...\hat{\underline
m}_5\hat{\underline\gamma}\hat{\underline\delta}}(\mu^{\hat{\underline\gamma}+}_Av^{\hat{\underline\delta}-}_A+\mu^{\hat{\underline\gamma}-}_{\dot
A}v^{\hat{\underline\delta}+}_{\dot A})=0
\end{equation}
that remove the contribution of 5-form coordinates
$z^{\hat{\underline\alpha}\hat{\underline\beta}}=z^{\hat{\underline
m}_1...\hat{\underline m}_5}\tilde\sigma_{\hat{\underline
m}_1...\hat{\underline
m}_5}{}^{\hat{\underline\alpha}\hat{\underline\beta}}$ associated
to tensorial central charge (TCC) generators $Z_{\hat{\underline
m}_1...\hat{\underline m}_5}$ of the $OSp(1|32)$ superalgebra.
Equation (\ref{x5}) involves inverse spinor harmonics
\begin{equation}
v^{\hat{\underline\alpha}}_{(\hat{\underline\alpha})}=(v^{\hat{\underline\alpha}-}_A,
v^{\hat{\underline\alpha}+}_{\dot A}):\quad
v^{\hat{\underline\alpha}}_{(\hat{\underline\alpha})}v_{\hat{\underline\alpha}}^{(\hat{\underline\beta})}=\delta_{(\hat{\underline\alpha})}^{(\hat{\underline\beta})}
\end{equation}
and the $OSp(1|32)$ invariant product of supertwistors is obtained
via the contraction with the orthosymplectic metric
\begin{equation}
G_{\mathbf{\Lambda\Sigma}}=\left(\begin{array}{ccc}
0& \delta_{\hat{\underline\alpha}}^{\hat{\underline\beta}}& 0\\[0.2cm]
-\delta_{\hat{\underline\beta}}^{\hat{\underline\alpha}}& 0& 0\\[0.2cm]
0 & 0 &  -i\\
\end{array}
\right).
\end{equation}

\section{Supertwistor form of $D=10$ superstring action}

Consider now the representation for $D=10$ $N=1$ superstring
action in terms of the above introduced $OSp(1|32)$ supertwistors.
Following the discussion in lower dimensions we start with the
$D=10$ Lorentz-harmonic superstring action \cite{BZstring}
\begin{equation}\label{action}
S_{LH}^{D=10}=S_{kin}+S_{WZ}
\end{equation}
with
\begin{equation}\label{kinterm}
S_{kin}=\frac{1}{2(\alpha')^{1/2}}\int
d^2\xi\left(e^{+2}n^{-2}_{\hat{\underline
m}}-e^{-2}n^{+2}_{\hat{\underline
m}}\right)\wedge\omega^{\hat{\underline m}}(d)+\frac{c}{2}\int
d^2\xi e^{-2}\wedge e^{+2},
\end{equation}
and
\begin{equation}
S_{WZ}=\frac{is}{c\alpha'}\int d^2\xi\omega^{\hat{\underline
m}}(d)\wedge
d\theta^{\hat{\underline\alpha}}\sigma_{\hat{\underline
m}\hat{\underline\alpha}\hat{\underline\beta}}\theta^{\hat{\underline\beta}},
\end{equation}
where $\omega^{\hat{\underline m}}(d)=dx^{\hat{\underline
m}}-id\theta^{\hat{\underline\alpha}}\sigma^{\hat{\underline
m}}_{\hat{\underline\alpha}\hat{\underline\beta}}\theta^{\hat{\underline\beta}}$
is the $D=10$ $N=1$ supersymmetric 1-form. The action
(\ref{kinterm}) also includes $n^{\pm2}_{\hat{\underline m}}(\xi)$
components of the string adopted $D=10$ vector harmonic matrix
$n_{\hat{\underline m}}^{(\hat{\underline
n})}=(n^{+2}_{\hat{\underline m}}, n^{-2}_{\hat{\underline m}},
n^I_{\hat{\underline m}})$
\begin{equation}\label{norm10}
\begin{array}{c}
n_{\hat{\underline m}}^{(\hat{\underline k})}n^{\hat{\underline
m}(\hat{\underline n})}=\eta^{(\hat{\underline k})(\hat{\underline n})}:\\[0,2cm]
n_{\hat{\underline m}}^{+2}n^{\hat{\underline
m}+2}=n_{\hat{\underline m}}^{-2}n^{\hat{\underline m}-2}=0,\quad
n_{\hat{\underline m}}^{+2}n^{\hat{\underline m}-2}=2,\quad
n_{\hat{\underline m}}^{\pm2}n^{\hat{\underline m}I}=0,\quad
n_{\hat{\underline m}}^{I}n^{\hat{\underline m}J}=-\delta^{IJ}
\end{array}
\end{equation}
that have the following representation in terms of the spinor
harmonics
\begin{equation}\label{vecspin10}
n^{+2}_{\hat{\underline
m}}={\textstyle\frac18}v^+_{\hat{\underline\alpha}
A}\tilde\sigma_{\hat{\underline
m}}^{\hat{\underline\alpha}\hat{\underline\beta}}v^+_{\hat{\underline\beta}
A},\quad n_{\hat{\underline
m}}^{-2}={\textstyle\frac18}v^-_{\hat{\underline\alpha}\dot
A}\tilde\sigma_{\hat{\underline
m}}^{\hat{\underline\alpha}\hat{\underline\beta}}v^-_{\hat{\underline\beta}\dot
A},\quad n_{\hat{\underline
m}}^{I}={\textstyle\frac18}v^+_{\hat{\underline\alpha}
A}\tilde\sigma_{\hat{\underline
m}}^{\hat{\underline\alpha}\hat{\underline\beta}}v^-_{\hat{\underline\beta}\dot
A}\gamma^{I}_{A\dot A},
\end{equation}
where $I,J=1,...,8$ stand for the $SO(8)$ vector indices and
$\gamma^{I}_{A\dot A}$ are the $8d$ analogs of the $4-$dimensional
$\sigma-$matrices. Their differentials consistent with the
constraints (\ref{norm10})
\begin{equation}
dn^{\pm2}_{\hat{\underline
m}}=\mp\frac12\hat\Omega^{+2-2}(d)n_{\hat{\underline
m}}^{\pm2}+\hat\Omega^{\pm2I}(d)n^{I}_{\hat{\underline m}},\quad
dn^{I}_{\hat{\underline
m}}=\frac12\hat\Omega^{+2I}(d)n^{-2}_{\hat{\underline
m}}+\frac12\hat\Omega^{-2I}(d)n^{+2}_{\hat{\underline
m}}+\hat\Omega^{IJ}(d)n_{\hat{\underline m}}^{J}
\end{equation}
depend on the $SO(1,1)\times SO(8)$ split components of the Cartan
1-form $\hat\Omega^{(\hat{\underline k})(\hat{\underline l})}=
\frac12(n_{\hat{\underline m}}^{(\hat{\underline
k})}dn^{\hat{\underline m}(\hat{\underline l})}-n_{\hat{\underline
m}}^{(\hat{\underline l})}dn^{\hat{\underline m}(\hat{\underline
k})})$:
\begin{equation}\label{cartan10}
\begin{array}{c}
\hat\Omega^{+2-2}(d)=\frac12(n^{+2}_{\hat{\underline
m}}dn^{\hat{\underline m}-2}-n^{-2}_{\hat{\underline
m}}dn^{\hat{\underline m}+2}),\quad
\hat\Omega^{\pm2I}(d)=\frac12(n_{\hat{\underline
m}}^{\pm2}dn^{\hat{\underline
m}I}-n_{\hat{\underline m}}^{I}dn^{\hat{\underline m}\pm2}),\\[0.2cm]
\hat\Omega^{IJ}(d)=\frac12(n_{\hat{\underline
m}}^{I}dn^{\hat{\underline m}J}-n_{\hat{\underline
m}}^{J}dn^{\hat{\underline m}I})
\end{array}
\end{equation}
invariant under $SO(1,9)$ Lorentz transformations acting on the
indices without brackets.

Similarly to the lower dimensional cases twistorization of the
action (\ref{action}) follows by expressing
$\omega^{\hat{\underline m}}(d)$ projections on the vector
harmonics (and using (\ref{vecspin10})) through the $D=10$
supertwistors (\ref{deftwist10})
\begin{equation}\label{bridge10}
\varpi^{+2}(d)=\omega^{\hat{\underline
m}}(d)n^{+2}_{\hat{\underline
m}}=\frac18dZ^{\mathbf\Lambda+}_AG_{\mathbf{\Lambda\Sigma}}Z^{\mathbf\Sigma+}_A,\quad\varpi^{-2}(d)=\omega^{\hat{\underline
m}}(d)n^{-2}_{\hat{\underline
m}}=\frac18dZ^{\mathbf\Lambda-}_{\dot
A}G_{\mathbf{\Lambda\Sigma}}Z^{\mathbf\Sigma-}_{\dot A},
\end{equation}
and
\begin{equation}\label{bridge10'}
\varpi^{I}(d)=\omega^{\hat{\underline m}}(d)n_{\hat{\underline
m}}^{I}=\frac{1}{16}\gamma^{I}_{A\dot
A}(dZ^{\mathbf\Lambda+}_AG_{\mathbf{\Lambda\Sigma}}Z^{\mathbf{\Sigma}-}_{\dot
A}+dZ^{\mathbf\Lambda-}_{\dot
A}G_{\mathbf{\Lambda\Sigma}}Z^{\mathbf\Sigma+}_A),
\end{equation}
provided $Z^{\mathbf\Lambda+}_A$ and $Z^{\mathbf\Sigma-}_{\dot A}$
obey the constraints (\ref{x3}), (\ref{x5}). So that the $D=10$
superstring action (\ref{action}) acquires the form
\begin{equation}\label{action-tw}
\begin{array}{c}
S^{D=10}_{tw}=\frac{1}{2(\alpha')^{1/2}}\int d^2\xi(e^{+2}\wedge\varpi^{-2}(d)-e^{-2}\wedge\varpi^{+2}(d))+\frac{c}{2}\int d^2\xi e^{-2}\wedge e^{+2}\\[0.2cm]
+\frac{is}{c\alpha'}\int
d^2\xi(\frac12\varpi^{-2}(d)\wedge\phi^{+2}(d)+\frac12\varpi^{+2}(d)\wedge\phi^{-2}(d)-\varpi^{I}(d)\wedge\phi^{I}(d)).
\end{array}
\end{equation}
In (\ref{action-tw}) $\phi(d)$ are the 1-forms quadratic in the
Grassmann-odd supertwistor components
\begin{equation}\label{phi10}
\phi^{+2}(d)=\frac18{\cal
D}\eta^+_A\eta^+_A,\quad\phi^{-2}(d)=\frac18{\cal D}\eta^-_{\dot
A}\eta^-_{\dot A},\quad\phi^{I}(d)=\frac{1}{16}\gamma^{I}_{A\dot
A}({\cal D}\eta^+_A\eta^-_{\dot A}+{\cal D}\eta^-_{\dot
A}\eta^+_A),
\end{equation}
and $SO(1,1)\times SO(8)$ covariant differentials of $\eta^+_A$
and $\eta^-_{\dot A}$ are defined as
\begin{equation}
\begin{array}{c}
{\cal D}\eta^+_A=d\eta^+_A+\frac14\hat\Omega^{+2-2}(d)\eta^+_A-\frac12\hat\Omega^{+2I}(d)\gamma^{I}_{A\dot A}\eta^-_{\dot A}-\frac14\hat\Omega^{IJ}(d)\gamma^{IJ}_{AB}\eta^{+}_B,\\[0.2cm]
{\cal D}\eta^-_{\dot A}=d\eta^-_{\dot
A}-\frac14\hat\Omega^{+2-2}(d)\eta^-_{\dot
A}-\frac12\hat\Omega^{-2I}(d)\tilde\gamma^{I}_{\dot
AA}\eta^+_{A}-\frac14\hat\Omega^{IJ}(d)\tilde\gamma^{IJ}_{\dot
A\dot B}\eta^{-}_{\dot B},
\end{array}
\end{equation}
where $\gamma^{IJ}_{AB}=\frac12(\gamma^I_{A\dot
A}\tilde\gamma^{J}_{\dot AB}-\gamma^J_{A\dot
A}\tilde\gamma^{I}_{\dot AB})$, $\tilde\gamma^{IJ}_{\dot A\dot
B}=\frac12(\tilde\gamma^I_{\dot AA}\gamma^{J}_{A\dot
B}-\tilde\gamma^J_{\dot AA}\gamma^{I}_{A\dot B})$ are the
$Spin(8)$ generators in the $s$ and $c$ representations.

Admissible differentials of the supertwistors (\ref{deftwist10})
that respect the constraints (\ref{x3}), (\ref{x5}) can be brought
to the form
\begin{equation}\label{diffsuper}
\begin{array}{rl}
dZ^{\mathbf\Lambda+}_A=&-\frac14\hat\Omega^{+2-2}(d)Z^{\mathbf\Lambda+}_A+\frac12\hat\Omega^{+2I}(d)\gamma^{I}_{A\dot A}Z^{\mathbf\Lambda-}_{\dot A}+\frac14\hat\Omega^{IJ}(d)\gamma^{IJ}_{AB}Z^{\mathbf\Lambda+}_B\\[0.2cm]
&+(\delta_{AB}\varpi^{+2}(d)+i{\cal
D}\eta^+_A\eta^+_B)V^{\mathbf\Lambda-}_B+(\gamma^{I}_{A\dot
B}\varpi^{I}(d)+i{\cal D}\eta^+_{A}\eta^-_{\dot B})V^{\mathbf\Lambda+}_{\dot B}+J^{\mathbf\Lambda}{}_{\mathbf\Sigma}{\cal D}Z^{\mathbf\Sigma+}_A;\\[0.2cm]
dZ^{\mathbf\Lambda-}_{\dot A}=&\frac14\hat\Omega^{+2-2}(d)Z^{\mathbf\Lambda-}_{\dot A}+\frac12\hat\Omega^{-2I}(d)\tilde\gamma^{I}_{\dot AA}Z^{\mathbf\Lambda+}_{A}+\frac14\hat\Omega^{IJ}(d)\tilde\gamma^{IJ}_{\dot A\dot B}Z^{\mathbf\Lambda-}_{\dot B}\\[0.2cm]
&+(\delta_{\dot A\dot B}\varpi^{-2}(d)+i{\cal D}\eta^-_{\dot
A}\eta^-_{\dot B})V^{\mathbf\Lambda+}_{\dot
B}+(\tilde\gamma^{I}_{\dot AB}\varpi^{I}(d)+i{\cal D}\eta^-_{\dot
A}\eta^+_{B})V^{\mathbf\Lambda-}_{B}+J^{\mathbf\Lambda}{}_{\mathbf\Sigma}{\cal
D}Z^{\mathbf\Sigma-}_{\dot A},
\end{array}
\end{equation}
where $V^{\mathbf\Lambda-}_A=(v^{\hat{\underline\alpha}-}_A,\ 0,\
0)$, $V^{\mathbf\Lambda+}_{\dot
A}=(v^{\hat{\underline\alpha}+}_{\dot A},\ 0,\ 0)$ and
\begin{equation}
J^{\mathbf\Lambda}{}_{\mathbf\Sigma}=\left(
\begin{array}{ccc}
0&0&0\\
0&0&0\\
0&0&1
\end{array}
\right).
\end{equation}
In deriving (\ref{diffsuper}) used were the expressions for the
admissible differentials of the spinor harmonics
\begin{equation}\label{sdiff}
\begin{array}{c}
dv^+_{\hat{\underline\alpha} A}=-\frac14\hat\Omega^{+2-2}(d)v^+_{\hat{\underline\alpha} A}+\frac12\hat\Omega^{+2I}(d)\gamma^{I}_{A\dot A}v^-_{\hat{\underline\alpha}\dot A}+\frac14\hat\Omega^{IJ}(d)\gamma^{IJ}_{AB}v^+_{\hat{\underline\alpha} B},\\[0.2cm]
dv^-_{\hat{\underline\alpha}\dot
A}=\frac14\hat\Omega^{+2-2}(d)v^-_{\hat{\underline\alpha}\dot
A}+\frac12\hat\Omega^{-2I}(d)\tilde\gamma^{I}_{\dot
AA}v^+_{\hat{\underline\alpha}
A}+\frac14\hat\Omega^{IJ}(d)\tilde\gamma^{IJ}_{\dot A\dot
B}v^-_{\hat{\underline\alpha}\dot B}.
\end{array}
\end{equation}
The derivation coefficients (\ref{cartan10}) can equivalently be
rewritten in terms of the spinor harmonics as
\begin{equation}\label{Cartanforms}
\begin{array}{c}
\hat\Omega^{+2-2}(d)=\frac14(dv^-_{\hat{\underline\alpha}\dot A}v^{\hat{\underline\alpha}+}_{\dot A}-dv^+_{\hat{\underline\alpha} A}v^{\hat{\underline\alpha}-}_A),\\[0.2cm]
\hat\Omega^{+2I}(d)=\frac14dv^+_{\hat{\underline\alpha} A}\gamma^{I}_{A\dot A}v^{\hat{\underline\alpha}+}_{\dot A},\ \hat\Omega^{-2I}(d)=\frac14dv^-_{\hat{\underline\alpha}\dot A}\tilde\gamma^{I}_{\dot AA}v^{\hat{\underline\alpha}-}_A,\\[0.2cm]
\hat\Omega^{IJ}(d)=\frac18(dv^+_{\hat{\underline\alpha}
A}\gamma^{IJ}_{AB}v^{\hat{\underline\alpha}-}_{B}+dv^-_{\hat{\underline\alpha}\dot
A}\tilde\gamma^{IJ}_{\dot A\dot
B}v^{\hat{\underline\alpha}+}_{\dot B}).
\end{array}
\end{equation}
Then the differentials of the 1-forms (\ref{bridge10}),
(\ref{bridge10'}), (\ref{phi10}) equal
\begin{equation}
\begin{array}{rl}
d\varpi^{+2}=&\frac12\hat\Omega^{+2-2}(d)\wedge\varpi^{+2}(d)-\hat\Omega^{+2I}(d)\wedge\varpi^{I}(d)-\frac{i}{8}{\cal D}\eta^+_A\wedge{\cal D}\eta^+_A,\\[0.2cm]
d\varpi^{-2}=&-\frac12\hat\Omega^{+2-2}(d)\wedge\varpi^{-2}(d)-\hat\Omega^{-2I}(d)\wedge\varpi^{I}(d)-\frac{i}{8}{\cal D}\eta^-_{\dot A}\wedge{\cal D}\eta^-_{\dot A},\\[0.2cm]
d\varpi^I=&-\frac12\hat\Omega^{+2I}(d)\wedge\varpi^{-2}(d)-\frac12\hat\Omega^{-2I}(d)\wedge\varpi^{+2}(d)-\hat\Omega^{IJ}(d)\wedge\varpi^J(d)\\[0.2cm]
-&\frac{i}{8}\mathcal D\eta^+_A\wedge\mathcal D\eta^-_{\dot
A}\gamma^I_{A\dot A}
\end{array}
\end{equation}
and
\begin{equation}
\begin{array}{rl}
d\phi^{+2}=&\frac12\hat\Omega^{+2-2}(d)\wedge\phi^{+2}(d)-\hat\Omega^{+2I}(d)\wedge\phi^{I}(d)+\frac18\mathcal D\eta^{+}_{A}\wedge\mathcal D\eta^{+}_{A},\\[0.2cm]
d\phi^{-2}=&-\frac12\hat\Omega^{+2-2}(d)\wedge\phi^{-2}(d)-\hat\Omega^{-2I}(d)\wedge\phi^{I}(d)+\frac{1}{8}\mathcal D\eta^{-}_{\dot A}\wedge\mathcal D\eta^{-}_{\dot A},\\[0.2cm]
d\phi^{I}=&-\frac12\hat\Omega^{+2I}(d)\wedge\phi^{-2}(d)-\frac12\hat\Omega^{-2I}(d)\wedge\phi^{+2}(d)-\hat\Omega^{IJ}(d)\wedge\phi^{J}(d)\\[0.2cm]
+&\frac{1}{8}\mathcal D\eta^{+}_{A}\wedge\mathcal D\eta^{-}_{\dot
A}\gamma^{I}_{A\dot A}.
\end{array}
\end{equation}
So that the action (\ref{action-tw}) produces the following
variation
\begin{equation}\label{var10}
\begin{array}{rl}
\delta S^{D=10}_{tw}=&\frac{1}{2(\alpha')^{1/2}}\int d^2\xi\left(e^{+2}\wedge\left[-\frac12\hat\Omega^{+2-2}(d)\varpi^{-2}(\delta)+\frac12\hat\Omega^{+2-2}(\delta)\varpi^{-2}(d)-\hat\Omega^{-2I}(d)\varpi^{I}(\delta)\right.\right.\\[0.3cm]
+&\left.\hat\Omega^{-2I}(\delta)\varpi^{I}(d)-\frac{i}{4}{\cal D}\eta^-_{\dot A}{\cal D}(\delta)\eta^-_{\dot A}\right]-de^{+2}\varpi^{-2}(\delta)+\delta e^{+2}\wedge\varpi^{-2}(d)\\[0.3cm]
-& e^{-2}\wedge\left[\frac12\hat\Omega^{+2-2}(d)\varpi^{+2}(\delta)-\frac12\hat\Omega^{+2-2}(\delta)\varpi^{+2}(d)-\hat\Omega^{+2I}(d)\varpi^{I}(\delta)\right.\\[0.3cm]
+&\left.\hat\Omega^{+2I}(\delta)\varpi^{I}(d)-\frac{i}{4}{\cal D}\eta^+_{A}{\cal D}(\delta)\eta^+_{A}\right]+de^{-2}\varpi^{+2}(\delta)-\delta e^{-2}\wedge\varpi^{+2}(d)\\[0.3cm]
+&\left. c(\alpha')^{1/2}(\delta e^{-2}\wedge
e^{+2}+e^{-2}\wedge\delta e^{+2})\right)\\[0.2cm]
+&\frac{is}{c\alpha'}\int d^2\xi\left(\varpi^{+2}(d)\wedge{\cal D}\eta^-_{\dot A}{\cal D}(\delta)\eta^-_{\dot A}+\frac12\varpi^{+2}(\delta){\cal D}\eta^-_{\dot A}\wedge{\cal D}\eta^-_{\dot A}\right.\\[0.3cm]
+&\varpi^{-2}(d)\wedge{\cal D}\eta^+_{A}{\cal D}(\delta)\eta^+_{A}+\frac12\varpi^{-2}(\delta){\cal D}\eta^+_{A}\wedge{\cal D}\eta^+_{A}\\[0.3cm]
-&\left. \left[\varpi^{I}(d)\wedge({\cal D}\eta^+_A{\cal
D}(\delta)\eta^-_{\dot A}+{\cal D}\eta^-_{\dot A}{\cal
D}(\delta)\eta^+_{A})+\varpi^{I}(\delta){\cal
D}\eta^+_A\wedge{\cal D}\eta^-_{\dot A}\right]\gamma^{I}_{A\dot
A}\right).
\end{array}
\end{equation}
Considering $\varpi^{\pm2}(\delta)$, $\varpi^{I}(\delta)$, ${\cal
D}(\delta)\eta^+_A$, ${\cal D}(\delta)\eta^-_{\dot A}$ together
with $\hat\Omega^{+2-2}(\delta)$, $\hat\Omega^{\pm2I}(\delta)$,
$\hat\Omega^{IJ}(\delta)$ as independent variations yields the
supertwistor representation of the $D=10$ $N=1$ superstring
equations of motion. There are present nondynamical equations
\begin{equation}\label{rheotr1}
\varpi^{\pm2}(d)=c(\alpha')^{1/2}e^{\pm2},\quad \varpi^{I}(d)=0
\end{equation}
that are the $D=10$ analogs of the superspace formulation
equations (\ref{rheotr6}). Note that equations of motion resulting
from the variation w.r.t. $\hat\Omega^{+2-2}(\delta)$ and
$\hat\Omega^{IJ}(\delta)$ trivialize that is the consequence of
the $SO(1,1)\times SO(8)$ gauge invariance of the action
(\ref{action}), (\ref{action-tw}). Other independent bosonic
equations of motion read
\begin{equation}\label{eomxi}
e^{-2}\wedge\hat\Omega^{+2I}(d)-e^{+2}\wedge\hat\Omega^{-2I}(d)-{\textstyle\frac{is}{4c(\alpha')^{1/2}}}{\cal
D}\eta^+_A\gamma^{I}_{A\dot A}\wedge{\cal D}\eta^-_{\dot A}=0.
\end{equation}
Taking into account above derived equations one is able to present
the equations of motion for the Grassmann-odd components of
supertwistors in the following form
\begin{equation}\label{eomf}
(1+s)e^{-2}\wedge{\cal D}\eta^+_A=0,\quad(1-s)e^{+2}\wedge{\cal
D}\eta^-_{\dot A}=0
\end{equation}
similarly to the $D=6$ superstring.

The supertwistor representation (\ref{action-tw}) of the $D=10$
Lorentz-harmonic superstring action (\ref{action}) is thus
invariant under the $\kappa-$symmetry transformations. When $s=1$
explicit transformation laws are
\begin{equation}
\begin{array}{c}
\delta_\kappa
Z^{\mathbf\Lambda+}_A=\frac12\hat\Omega^{+2I}(\delta_\kappa)\gamma^{I}_{A\dot
A}Z^{\mathbf\Lambda-}_{\dot A},\\[0.2cm]
\delta_\kappa Z^{\mathbf\Lambda-}_{\dot
A}=\frac12\hat\Omega^{-2I}(\delta_\kappa)\tilde\gamma^{I}_{\dot
AA}Z^{\mathbf\Lambda+}_A-(K^{\mathbf\Sigma-}_{\dot
A}Z^-_{\mathbf\Sigma\dot B})V^{\mathbf\Lambda+}_{\dot B}
-(K^{\mathbf\Sigma-}_{\dot A}Z^+_{\mathbf\Sigma B})V^{\mathbf\Lambda-}_B+K^{\mathbf\Lambda-}_{\dot A},\\[0.3cm]
\delta_\kappa e^{+2}=0,\quad\delta_\kappa
e^{-2}=\frac{1}{4c(\alpha')^{1/2}}{\cal
D}Z^{\mathbf\Lambda-}_{\dot A}K^-_{\mathbf\Lambda\dot A},
\end{array}
\end{equation}
where
\begin{equation}
\hat\Omega^{\pm2I}(\delta_\kappa)=\pm\frac{1}{8c(\alpha')^{1/2}}e^{\mu\pm2}{\cal
D}_\mu Z^{\mathbf\Lambda+}_AK^-_{\mathbf\Lambda\dot
A}\gamma^{I}_{A\dot A}
\end{equation}
and $K^{\mathbf\Lambda-}_{\dot A}=(0, 0, \kappa^-_{\dot A})$ is
the supertwistor realization of the gauge parameter
$\kappa^-_{\dot A}(\xi)$. Correspondingly when $s=-1$ we obtain
\begin{equation}
\begin{array}{c}
\delta_\kappa
Z^{\mathbf\Lambda+}_A=\frac12\hat\Omega^{+2I}(\delta_\kappa)\gamma^{I}_{A\dot
A}Z^{\mathbf\Lambda-}_{\dot
A}-(K^{\mathbf\Sigma+}_{A}Z^+_{\mathbf\Sigma
B})V^{\mathbf\Lambda-}_{B}-(K^{\mathbf\Sigma+}_{A}Z^-_{\mathbf\Sigma\dot
B})V^{\mathbf\Lambda+}_{\dot B}+K^{\mathbf\Lambda+}_A,\\[0.2cm]
\delta_\kappa Z^{\mathbf\Lambda-}_{\dot
A}=\frac12\hat\Omega^{-2I}(\delta_\kappa)\tilde\gamma^{I}_{\dot
AA}Z^{\mathbf\Lambda+}_A,\\[0.3cm]
\delta_\kappa e^{+2}=\frac{1}{4c(\alpha')^{1/2}}{\cal
D}Z^{\mathbf\Lambda+}_{A}K^+_{\mathbf\Lambda A},\quad\delta_\kappa
e^{-2}=0,
\end{array}
\end{equation}
where
\begin{equation}
\hat\Omega^{\pm2I}(\delta_\kappa)=\pm\frac{1}{8c(\alpha')^{1/2}}e^{\mu\pm2}K^{\mathbf\Lambda+}_A{\cal
D}_\mu Z^-_{\mathbf\Lambda\dot A}\gamma^{I}_{A\dot A}
\end{equation}
with $K^{\mathbf\Lambda+}_{A}=(0, 0, \kappa^+_{A})$.

Quite analogously to the $D=4$ and $D=6$ cases the supertwistor
action (\ref{action-tw}) can be simplified if the gauge freedom
related to $\kappa-$symmetry is fixed. The gauge fixed action form
depends on the value of $s$ and is given by the expressions
\begin{equation}\label{gf10+1}
\begin{array}{rl}
S^{D=10}_{gf,s=1}=&{\textstyle\frac{1}{16(\alpha')^{1/2}}}\int
d^2\xi\left(e^{+2}\wedge dz^{\mathcal A-}_{\dot A}z^-_{\mathcal
A\dot
A}-e^{-2}\wedge d\tilde{Z}^{\mathbf\Lambda+}_{A}\tilde{Z}^+_{\mathbf\Lambda A}\right)\\[0.2cm]
+&\frac{c}{2}\int d^2\xi e^{-2}\wedge e^{+2}
\end{array}
\end{equation}
or
\begin{equation}\label{gf10-1}
\begin{array}{rl}
S^{D=10}_{gf,s=-1}=&{\textstyle\frac{1}{16(\alpha')^{1/2}}}\int
d^2\xi\left( e^{+2}\wedge d\tilde{Z}^{\mathbf\Lambda-}_{\dot
A}\tilde{Z}_{\mathbf\Lambda\dot A}^{-}-e^{-2}\wedge d
z^{\mathcal A+}_{A}z_{\mathcal A A}^{+}\right)\\[0.2cm]
+&\frac{c}{2}\int d^2\xi e^{-2}\wedge e^{+2}.
\end{array}
\end{equation}
In (\ref{gf10+1}), (\ref{gf10-1}) $z^{\mathcal A+}_A$,
$z^{\mathcal A-}_{\dot A}$ stand for bosonic $Sp(32)$ twistors and
there has been assumed the following redefinition of the
supertwistor components
\begin{equation}\label{tilde10}
\begin{array}{c}
\tilde{Z}^{\mathbf\Lambda+}_{A}=(\tilde\mu^{\hat{\underline\alpha}+}_{A},
v^{+}_{\hat{\underline\alpha} A}, \tilde\eta^{+}_{A}):\quad
\tilde\mu^{\hat{\underline\alpha}+}_{A}=(x^{\hat{\underline\alpha}\hat{\underline\beta}}-16i\theta^{\hat{\underline\alpha}}\theta^{\hat{\underline\beta}})v_{\hat{\underline\beta} A}^{+},\quad\tilde\eta^{+}_{A}=8v_{\hat{\underline\alpha} A}^{+}\theta^{\hat{\underline\alpha}},\\
\tilde{Z}^{\mathbf\Lambda-}_{\dot
A}=(\tilde\mu^{\hat{\underline\alpha}-}_{\dot A},
v^{-}_{\hat{\underline\alpha}\dot A}, \tilde\eta^{-}_{\dot
A}):\quad\tilde\mu^{\hat{\underline\alpha}-}_{\dot
A}=(x^{\hat{\underline\alpha}\hat{\underline\beta}}-16i\theta^{\hat{\underline\alpha}}\theta^{\hat{\underline\beta}})v_{\hat{\underline\beta}\dot
A}^{-},\quad\tilde\eta^{-}_{\dot A}=8v_{\hat{\underline\alpha}\dot
A}^{-}\theta^{\hat{\underline\alpha}}
\end{array}
\end{equation}
and the $OSp(32|1)$ metric
\begin{equation}\label{tildeg}
\tilde G_{\mathbf{\Lambda\Sigma}}=\left(\begin{array}{ccc}
0& \delta_{\hat{\underline\alpha}}^{\hat{\underline\beta}} & 0\\[0.2cm]
-\delta^{\hat{\underline\alpha}}_{\hat{\underline\beta}} & 0& 0\\[0.2cm]
0 & 0 & -\frac{i}{2}
\end{array}\right).
\end{equation}
Note that the above (super)twistors still have to satisfy the
constraints analogous to (\ref{x3}) and (\ref{x5}) to be incident
to $D=10$ vector coordinates $x^{\hat{\underline m}}$.

Adduced consideration of the twistor transform for $D=10$ $N=1$
superstring in the Lorentz-harmonic formulation can be generalized
to include Type II case as well. In particular, for the Type IIB
superstring choosing the $\kappa-$symmetry gauge conditions
$\eta^{1+}_A=\eta^{2+}_A$, $\eta^{1-}_{\dot A}=\eta^{2-}_{\dot A}$
results in the action functional with quadratic dependence on the
$N=1$ supertwistor variables (\ref{tilde10})
\begin{equation}\label{twis2b}
\begin{array}{rl}
S^{D=10,IIB}_{gf}=&{\textstyle\frac{1}{16(\alpha)^{1/2}}}\int
d^2\xi\left( e^{+2}\wedge d\tilde Z^{\mathbf\Lambda-}_{\dot
A}\tilde G_{\mathbf{\Lambda\Sigma}}\tilde Z^{\mathbf\Sigma-}_{\dot
A}-e^{-2}\wedge d\tilde Z^{\mathbf\Lambda+}_{A}\tilde
G_{\mathbf{\Lambda\Sigma}}\tilde
Z^{\mathbf\Sigma+}_{A}\right)\\[0.2cm]
+&{\textstyle\frac{c}{2}}\int d^2\xi e^{-2}\wedge e^{+2}.
\end{array}
\end{equation}

In conclusion let us discuss the relation between proposed
supertwistor formulation of the $D=10$ Lorentz-harmonic
superstrings and the light-cone gauge formulation of the
Green-Schwarz superstrings on example of the twistor transformed
Type IIB superstring action (\ref{twis2b}). To this end it is
convenient to introduce Lorentz-harmonics normalized up to a scale
factor
\begin{equation}
v^{\hat{\underline\alpha}}_{(\hat{\underline\alpha})}v^{(\hat{\underline\beta})}_{\hat{\underline\alpha}}=n\delta^{(\hat{\underline\beta})}_{(\hat{\underline\alpha})},\quad
n^{(\hat{\underline n})}_{\hat{\underline m}}n^{\hat{\underline
m}(\hat{\underline k})}=n^2\eta^{(\hat{\underline
n})(\hat{\underline k})}.
\end{equation}
Corresponding modification of the action (\ref{twis2b}) affects
only the last term
\begin{equation}
\frac{c}{2}\int d^2\xi e^{-2}\wedge e^{+2}\rightarrow
\frac{c}{2}\int d^2\xi n^2 e^{-2}\wedge e^{+2}.
\end{equation}
Its advantage is that the action functional becomes invariant
under the gauge Weyl scalings
\begin{equation}
e^{\pm2}_\mu\to e^{-\lambda}e^{\pm2}_\mu,\quad\tilde
Z^{\mathbf\Lambda+}_A\to e^{\lambda/2}\tilde
Z^{\mathbf\Lambda+}_A,\quad \tilde Z^{\mathbf\Lambda-}_{\dot A}\to
e^{\lambda/2}\tilde Z^{\mathbf\Lambda-}_{\dot A}
\end{equation}
that allows to gauge away all the zweibein components e.g. as
\begin{equation}\label{confg}
e^f_\mu=(\alpha')^{1/2}\delta^f_\mu,\quad
e^\mu_f=(\alpha')^{-1/2}\delta^\mu_f.
\end{equation}

Then consider the expansion of the supertwistor primary spinor
components over the basis of inverse spinor harmonics
\begin{equation}\label{expansion}
\begin{array}{c}
\mu^{\hat{\underline\alpha}+}_A=\frac{1}{n}\left[(x^{+2}\delta_{AB}+\frac{i}{4}\tilde\eta^+_A\tilde\eta^+_B)v^{\hat{\underline\alpha}-}_B+(x^I\gamma^I_{A\dot
B}+\frac{i}{4}\tilde\eta^+_A\tilde\eta^-_{\dot
B})v^{\hat{\underline\alpha}+}_{\dot
B}\right],\\[0.2cm]
\mu^{\hat{\underline\alpha}-}_{\dot
A}=\frac{1}{n}\left[(x^{-2}\delta_{\dot A\dot
B}+\frac{i}{4}\tilde\eta^-_{\dot A}\tilde\eta^-_{\dot
B})v^{\hat{\underline\alpha}+}_{\dot B}+(x^I\tilde\gamma^I_{\dot
AB}+\frac{i}{4}\tilde\eta^-_{\dot
A}\tilde\eta^+_B)v^{\hat{\underline\alpha}-}_B\right],
\end{array}
\end{equation}
where
\begin{equation}
x^{\pm2}=x^{\hat{\underline m}}n^{\pm2}_{\hat{\underline m}},\quad
x^I=x^{\hat{\underline m}}n^{I}_{\hat{\underline m}}.
\end{equation}
Expansion (\ref{expansion}) is used to represent quadratic in the
supertwistors 1-forms that enter the action (\ref{twis2b}) as
\begin{equation}
\begin{array}{rl}
{\textstyle\frac18}d\tilde Z^{\mathbf\Lambda-}_{\dot A}\tilde
G_{\mathbf{\Lambda\Sigma}}\tilde Z^{\mathbf\Sigma-}_{\dot A}=&
dx^{-2}-(dn^2+{\textstyle\frac12}\hat\Omega^{+2-2}(d))x^{-2}-\hat\Omega^{-2I}(d)(x^I+{\textstyle\frac{i}{32}}\tilde\eta^-\tilde\gamma^I\tilde\eta^+)\\[0.2cm]
-&{\textstyle\frac{i}{16}}d\tilde\eta^-_{\dot A}\tilde\eta^-_{\dot
A}-{\textstyle\frac{i}{64}}\hat\Omega^{IJ}(d)\tilde\eta^-\tilde\gamma^{IJ}\tilde\eta^-,
\end{array}
\end{equation}
\begin{equation}
\begin{array}{rl}
{\textstyle\frac18}d\tilde Z^{\mathbf\Lambda+}_{A}\tilde
G_{\mathbf{\Lambda\Sigma}}\tilde Z^{\mathbf\Sigma+}_{A}=&
dx^{+2}-(dn^2-{\textstyle\frac12}\hat\Omega^{+2-2}(d))x^{+2}-\hat\Omega^{+2I}(d)(x^I+{\textstyle\frac{i}{32}}\tilde\eta^+\gamma^I\tilde\eta^-)\\[0.2cm]
-&{\textstyle\frac{i}{16}}d\tilde\eta^+_{A}\tilde\eta^+_{A}-{\textstyle\frac{i}{64}}\hat\Omega^{IJ}(d)\tilde\eta^+\gamma^{IJ}\tilde\eta^+.
\end{array}
\end{equation}
So that the action (\ref{twis2b}) acquires the form
\begin{equation}\label{cartan2b}
\begin{array}{rl}
S^{D=10,IIB}_{gf}=&{\textstyle\frac{1}{2(\alpha)^{1/2}}}\int
d^2\xi \left[
e^{+2}\wedge(dx^{-2}-(dn^2+\frac12\hat\Omega^{+2-2}(d))x^{-2}-\hat\Omega^{-2I}(d)(x^I+\frac{i}{32}\tilde\eta^-\tilde\gamma^I\tilde\eta^+)\right.\\[0.2cm]
-&\left.\frac{i}{16}d\tilde\eta^-_{\dot A}\tilde\eta^-_{\dot A}-\frac{i}{64}\hat\Omega^{IJ}(d)\tilde\eta^-\tilde\gamma^{IJ}\tilde\eta^-)\right]\\[0.2cm]
-&{\textstyle\frac{1}{2(\alpha)^{1/2}}}\int d^2\xi \left[
e^{-2}\wedge(dx^{+2}-(dn^2-\frac12\hat\Omega^{+2-2}(d))x^{+2}-\hat\Omega^{+2I}(d)(x^I+\frac{i}{32}\tilde\eta^+\gamma^I\tilde\eta^-)\right.\\[0.2cm]
-&\left.\frac{i}{16}d\tilde\eta^+_{A}\tilde\eta^+_{A}-\frac{i}{64}\hat\Omega^{IJ}(d)\tilde\eta^+\gamma^{IJ}\tilde\eta^+)\right]+{\textstyle\frac{c}{2}}
\int d^2\xi n^2 e^{-2}\wedge e^{+2}
\end{array}
\end{equation}
depending on the components of the Cartan form (\ref{Cartanforms})
and $D=10$ $N=1$ superspace coordinates contracted with the vector
harmonics \footnote{Similar form of the $\kappa-$symmetry
gauge-fixed Lorentz-harmonic superstring action was previously
considered in \cite{U00}.}.

Consider now the representation for the Cartan 1-form components
defined by the pair of unconstrained 8-vectors $p^{\pm2I}$
\begin{equation}\label{anzatz}
\begin{array}{c}
\hat\Omega^{+2-2}(d)=2(dp^{+2I}p^{-2I}-p^{+2I}dp^{-2I}),\quad\hat\Omega^{\pm2I}(d)=2dp^{\pm2I},\\[0.2cm]
\hat\Omega^{IJ}(d)=p^{+2I}dp^{-2J}-dp^{-2I}p^{+2J}-dp^{+2I}p^{-2J}+p^{-2I}dp^{+2J}
\end{array}
\end{equation}
that up to quadratic terms satisfy Maurer-Cartan equations
\begin{equation}
\begin{array}{c}
d\hat\Omega^{+2-2}-\hat\Omega^{+2I}\wedge\hat\Omega^{-2I}=0,\quad
d\hat\Omega^{\pm2I}\mp{\textstyle\frac12}\hat\Omega^{+2-2}\wedge\hat\Omega^{\pm2I}+\hat\Omega^{\pm2J}\wedge\hat\Omega^{JI}=0,\\[0.2cm]
d\hat\Omega^{IJ}+{\textstyle\frac12}(\hat\Omega^{+2I}\wedge\hat\Omega^{-2J}+\hat\Omega^{-2I}\wedge\hat\Omega^{+2J})+\hat\Omega^{IK}\wedge\hat\Omega^{KJ}=0.
\end{array}
\end{equation}
Upon substituting (\ref{anzatz}) into (\ref{cartan2b}), choosing
the conformal gauge for the zweibein (\ref{confg}) and
concentrating on the quadratic terms one arrives at the following
action
\begin{equation}\label{prelc}
S'^{IIB}_{l.c.}=-{\textstyle\frac{2}{\alpha'}\int}
d^2\xi(p^{-2I}\partial_{-2}x^I+p^{+2I}\partial_{+2}x^I-2cp^{+2I}p^{-2I})+{\textstyle\frac{i}{16\alpha'}\int}
d^2\xi(\partial_{-2}\tilde\eta^-_{\dot A}\tilde\eta^-_{\dot
A}+\partial_{+2}\tilde\eta^+_A\tilde\eta^+_A)
\end{equation}
that has to be supplemented by the Virasoro constraints
\begin{equation}
\partial_{+2}x^{-2}+2p^{-2I}\partial_{+2}x^I-{\textstyle\frac{i}{16}}\partial_{+2}\tilde\eta^-_{\dot A}\tilde\eta^-_{\dot A}=0,\quad
\partial_{-2}x^{+2}+2p^{+2I}\partial_{-2}x^I-{\textstyle\frac{i}{16}}\partial_{-2}\tilde\eta^+_{A}\tilde\eta^+_{A}=0
\end{equation}
that determine $x^{\pm2}$ variables. It should be noted that the
degrees of freedom counting for the action (\ref{prelc}) precisely
matches that for the supertwistor action (\ref{twis2b}) it was
derived from provided the constraints on harmonics and
supertwistor variables are taken into account. Varying
(\ref{prelc}) w.r.t. $p^{\pm2I}$ yields
\begin{equation}\label{pi2}
p^{\pm2I}={\textstyle\frac{1}{2c}}\partial_{\mp2}x^I.
\end{equation}
So that $p^{\pm2I}$ admit interpretation of the linear
combinations of momenta conjugate to $\partial_\tau x^I$ and
$\partial_\sigma x^I$. Substituting (\ref{pi2}) back into
(\ref{prelc}) gives the light-cone gauge action for the IIB
superstring
\begin{equation}
S^{IIB}_{l.c.}=-{\textstyle\frac{1}{c\alpha'}\int}
d^2\xi(\partial_{+2}x^I\partial_{-2}x^I+{\textstyle\frac{i}{16}}\tilde\eta^-_{\dot
A}\partial_{-2}\tilde\eta^-_{\dot
A}+{\textstyle\frac{i}{16}}\tilde\eta^+_A\partial_{+2}\tilde\eta^+_A).
\end{equation}

\section{Conclusion}

There has been considered the reformulation in terms of
supertwistor variables of the first-order Lorentz-harmonic
superstring action in 6 and 10 dimensions. Relevant
higher-dimensional supertwistors realize the fundamental
representations of the respective minimal superconformal groups
and include spinor Lorentz-harmonics as their projectional parts
to match with the string momentum density representation, and
satisfy the set of algebraic constraints, whose solution can be
cast into the form of higher-dimensional generalization of the
Penrose-Ferber incidence relations with the conventional
superspace coordinates. Obtained have been the expressions for the
admissible variations/differentials of the supertwistor variables
that respect these constraints. The derivation coefficients then
appear as the independent parameters in the action functional
variation yielding the set of superstring equations of motion in
the supertwistor form. Analogously to considered earlier twistor
transformed $D=4$ superstring, proposed here supertwistor
formulation for the superstring in $D=6,10$ dimensions is
invariant under the irreducible $\kappa-$symmetry. Corresponding
transformation rules for supertwistors and auxiliary variables
have been explicitly given. Analyzed has been the possibility of
simplification of the supertwistor formulation by covariantly
gauge fixing $\kappa-$symmetry leading to the quadratic
superstring action that is of the 2-dimensional free field theory
type modulo the constraints on supertwistors. It was also shown
that by fixing remaining gauge freedom the action can be reduced
to that for the Green-Schwarz superstring in the light-cone gauge.
Though we have started with the Lorentz-harmonic superstring
action that is classically equivalent to the Green-Schwarz one,
and in the appropriate gauge proposed supertwistor formulation
matches the light-cone gauge superstring action, the issue of
quantization of the considered supertwistor representation for the
superstring awaits its solution.

\section{Acknowledgements}
Author is indebted to A.A.~Zheltukhin for valuable discussions and
the Abdus Salam ICTP, were part of this work was done, for the
warm hospitality.

 \end{document}